\documentclass[fleqn,usenatbib]{mnras}

\usepackage{newtxtext,newtxmath}

\usepackage[T1]{fontenc}

\DeclareRobustCommand{\VAN}[3]{#2}
\let\VANthebibliography\thebibliography
\def\thebibliography{\DeclareRobustCommand{\VAN}[3]{##3}\VANthebibliography}

\usepackage{graphicx}	
\usepackage{amsmath}	
\usepackage{mathtools}
\usepackage{amssymb}	
\usepackage{siunitx}    
\usepackage{xspace}     
\usepackage[capitalize]{cleveref}   
\usepackage{booktabs}   
\usepackage{mhchem}
\usepackage{threeparttable}

\newcommand{\corot}{{\sc CoRoT}\xspace}

\newcommand{\tess}{{\sc TESS}\xspace}
\DeclareSIUnit{\dex}{dex}
\DeclareSIUnit{\arcsecond}{as}
\DeclareSIUnit{\solarmass}{\ensuremath{M_\odot}\xspace}
\DeclareSIUnit{\solarradius}{\ensuremath{R_\odot}\xspace}
\DeclareSIUnit{\solarlum}{\ensuremath{L_\odot}\xspace}
\DeclareSIUnit{\year}{yr}
\newcommand{\var}[2]{\ensuremath{#1_{\textup{#2}}\,}\xspace}
\newcommand{\kepl}{\mbox{Kepler-444}\xspace}
\newcommand{\dnu}{\ensuremath{\Delta\nu}\xspace}
\newcommand{\numax}{\ensuremath{\nu_\textup{max}}\xspace}
\newcommand{\feh}{\ensuremath{[\textup{Fe}/\textup{H}]}\xspace}

\newcommand{\teff}{\ensuremath{\var{T}{eff}}\xspace}
\newcommand{\logg}{\ensuremath{\log g}\xspace}
\newcommand{\lphot}{\ensuremath{\var{L}{Phot}}\xspace}
\newcommand{\gc}{\ensuremath{g_\textup{cut}}\xspace}
\newcommand{\tcore}{\ensuremath{\tau_\mathrm{core}}\xspace}
\newcommand{\buld}{{\rm B19}\xspace}
\newcommand{\bm}[1]{\ensuremath{\mathcal{M}_{#1}}\xspace}
\newcommand{\basta}{{\tt BASTA}\xspace}
\newcommand{\kepler}{\textit{Kepler}\xspace}

\newcommand{\norat}{\ensuremath{\notin_{01}^{21}}\xspace}

\usepackage{transparent}
\newcommand{\ts}{\rule{0pt}{2.6ex}}

\title[The Convective Core of Kepler-444]{Did Kepler-444 have a long-lived convective core?}

\author[M. L. Winther et al.]{Mark Lykke Winther$^{1}$\thanks{E-mail: mark@phys.au.dk},
Víctor Aguirre B\o rsen-Koch$^{2}$\thanks{Formerly Víctor Silva Aguirre},
Jakob Lysgaard Rørsted$^{1}$\thanks{Formerly Jakob Rørsted Mosumgaard},\newauthor
Amalie Stokholm$^{3,4,1}$,
and Kuldeep Verma$^{5,1}$.
\\
$^1$Stellar Astrophysics Centre, Department of Physics and Astronomy, Aarhus University, Ny Munkegade 120, DK-8000 Aarhus C, Denmark\\
$^2$DARK, Niels Bohr Institute, University of Copenhagen, Jagtvej 128, 2200 Copenhagen, Denmark\\
$^3$Dipartimento di Fisica e Astronomia, Universit\`{a} degli Studi di Bologna, Via Gobetti 93/2, I-40129 Bologna, Italy\\
$^4$INAF -- Osservatorio di Astrofisica e Scienza dello Spazio di Bologna, Via Gobetti 93/3, I-40129 Bologna, Italy\\
$^5$Department of Physics, Indian Institute of Technology (BHU), Varanasi-221005, India
}

\date{Accepted XXX. Received YYY; in original form ZZZ}

\pubyear{2023}

\begin{document}
\label{firstpage}
\pagerange{\pageref{firstpage}--\pageref{lastpage}}
\maketitle

\begin{abstract}
With the greater power to infer the state of stellar interiors provided by asteroseismology, it has become possible to study the survival of initially convective cores within stars during their main-sequence evolution. Standard theories of stellar evolution predict that convective cores in sub-solar mass stars have lifetimes below $1\,{\rm Gyr}$. However, a recent asteroseismic study of the star Kepler-444 concluded that the initial convective core had survived for nearly $8\,{\rm Gyr}$. The goal of this paper is to study the convective-core evolution of Kepler-444 and to investigate its proposed longevity. We modify the input physics of stellar models to induce longer convective-core lifetimes and vary the associated parameter across a dense grid of evolutionary tracks. The observations of metallicity, effective temperature, mean density, and asteroseismic frequency ratios are fitted to the models using the {\tt BASTA} pipeline. We explore different choices of constraints, from which a long convective-core lifetime is only recovered for a few specific combinations: mainly from the inclusion of potentially unreliable frequencies and/or excluding the covariances between the frequency ratios; while for the classical parameters, the derived luminosity has the largest influence. For all choices of observables, our analysis reliably constrains the convective-core lifetime of Kepler-444 to be short, with a median around $0.6\,{\rm Gyr}$ and a $1\sigma$ upper bound around $3.5\,{\rm Gyr}$.
\end{abstract}

\begin{keywords}
asteroseismology -- convection -- stars: evolution -- stars: interiors -- stars: oscillations -- stars: low-mass
\end{keywords}



\section{Introduction}
\label{sec:introduction}
Understanding and accurately modelling stars is essential to understanding the formation and evolution of the Universe. For instance, accurate determinations of stellar ages and chemical abundances, which are necessary for that, rely on matching observations to theoretical models. While only the surface of stars can be observed, the interior structure and evolution have a major influence on the fundamental stellar parameters such as its age. However, thanks to the \corot, \kepler, and \tess missions \citep{baglin2006,Kepler1,Kepler2,ricker2014}, it is now possible to measure the intrinsic global oscillations in solar-like stars. The study of these is called \emph{asteroseismology} and it concerns how these oscillations can be used to prescribe and constrain the internal structure of the stars \citep[e.g.][]{Chaplin_Miglio_2013}. As these oscillations probe different depths into the stellar interior, these make it possible to better study stellar structure and interior processes. This is especially important in the study of convection within stars as it is currently one of the least understood phenomena that simultaneously has a large impact on the evolution \citep[see i.e.][]{Nordlund2009,Trampedach2010,SalarisCassisi2017}. Asteroseismology provides the means to study and constrain the appearance of convective zones in stellar interiors and thus allows us to adjust and improve our stellar models to provide more accurate descriptions of stars.

A variety of asteroseismic quantities can be determined depending on the quality of observations. The two most readily available are the mean large frequency separation, $\Delta\nu$, and the frequency of maximum power, $\nu_{\rm max}$, that can be used to match observations to models like classical parameters of stars such as the effective temperature and metallicity. Given a higher quality of the data, the individual frequencies can be determined. These are denoted by $\nu_{n,\ell}$, where $n$ is the radial order and $\ell$ the spherical degree. The observed individual frequencies provide better constraints when comparing them to the corresponding model frequencies, however such comparisons are subject to the problem of the so-called asteroseismic surface effect \citep{Brown1984,ChristensenDalsgaard1984,ChristensenDalsgaard1988}. This problem is the result of a model effect, where a simple parametrisation of highly turbulent convection using mixing-length theory leads to substantial errors in the predicted stratification of the outermost layers and hence in the calculated adiabatic oscillation frequencies. Furthermore, it is also caused by the non-adiabatic effects and the uncertainty in our understanding of the convection-oscillation interactions which makes the model frequencies uncertain. This overall causes there to be a deviation between model and observed frequencies, increasing with higher frequencies. It is therefore necessary to account for this by applying an ad-hoc surface correction \citep{Kjeldsen2008,BG14,Sonoi2015}, which relies on additional parameters, when comparing individual frequencies.

A second option is to circumvent the surface effect by instead using the \textit{frequency separation ratios}, or simply called the frequency ratios. Proposed by \citet{Roxburgh2003,Roxburgh2013}, these are constructed to be independent of the surface effect and instead be sensitive to the interior structure/convective zones of the star. This allows for the determination of e.g.\@ convective core size and convective overshoot \citep{SilvaAguirre2011,SilvaAguirre2013,Deheuvels2016,Zhang2022}. The frequency ratios are constructed as
\begin{align}
\begin{split}
    r_{01}\left(n\right) &= \frac{d_{01}\left(n\right)}{\dnu_1 \left(n\right)}\,, \\
    r_{02}\left(n\right) &= \frac{d_{0,2}\left(n\right)}{\dnu_1 \left(n\right)} \, ,
\end{split}
\label{eq:ratios}
\end{align}
with the individual large frequency separations defined as $\Delta\nu_l\left(n\right)=\nu_{n,l} - \nu_{n-1,l}$. The five-point small frequency separations are given as 
\begin{align}
    d_{01} \left(n\right) &= \frac{1}{8} \left(\nu_{n-1,0} - 4\nu_{n-1,1} + 6\nu_{n,0} - 4\nu_{n,1} + \nu_{n+1,0} \right) \,,
\label{eq:five_point}
\end{align}
and the small frequency separation 
\begin{align}
d_{0,2}\left(n\right)= \nu_{n,0} - \nu_{n-1,2}\,.
\label{eq:d02}
\end{align}
An alternative formulation of the ratios is prescribed using the three-point small frequency separation
\begin{align}
    d^*_{01} = 2\nu_{n,0} - \left(\nu_{n-1,1} + \nu_{n,1} \right)/2\,,
    \label{eq:three_point}
\end{align}
which is less correlated than the five-point separation but also less smooth \citep{Roxburgh2009}. The five-point definition is used throughout this work unless otherwise stated. Similar prescriptions can be used to determined the $r_{10}(n)$ ratios, but as it is constructed from the same frequencies as $r_{01}(n)$, using both would lead to overfitting of the data \citep{Roxburgh2018} and thus $r_{10}(n)$ is not used in this work. Instead, the ratio sequences $r_{01} = \left\{r_{01}(n), r_{01}(n+1),\dots\right\}$ and $r_{02} = \left\{r_{02}(n), r_{02}(n+1),\dots\right\}$ can be combined into the $r_{012}$ ratio sequence and used in the fitting scheme to match observations to models \citep[as e.g.][]{Verma2022}.

\kepl (also known as KIC 6278762, HIP 94931, and KOI-3158) is one such star with observations of adequate quality to extract individual frequencies. As it is a bright star, it has traditionally been in many studies and surveys \citep{Roman1955,Eggen1956,Wilson1962,Nordstrom2004,vanLeeuwen2007}. Additionally, since the first transit-like signals exoplanets were detected \citep{Tenenbaum2013}, its compact system of five exoplanets have been extensively studied \citep{Akeson2013,Campante2015,Dupuy2016,Bourrier2017,Pezzotti2021,Stalport2022}. More recently, the star itself has been the subject of several asteroseismic studies, partly also to improve the determination of the exoplanet properties. \cite{Campante2015} used observations from \kepler to validate its 5 exoplanets through transit analysis. They derived $\Delta\nu$ and $\nu_{\rm max}$ from the observations and used them alongside classical parameters determined from spectroscopy to deduce the fundamental stellar properties of the star using various pipelines. \cite{SilvaAguirre2015} were able to use individual oscillation frequencies of an ensemble of \kepler stars including \kepl, to determine fundamental stellar properties using a Bayesian scheme. Both studies determined the star to be an old $(\SI{\sim11\pm 1}{Gyr})$ sub-solar mass $(\SI{\sim0.748\pm0.044}{\solarmass})$ star.

A recent study by \citet[][hereafter \buld]{Buldgen2019} re-analysed the star and found the same fundamental parameters. They also investigated the evolution of its convective core, which was not previously determined in the other studies. From detailed forward modelling using the observed frequency ratios, it was determined that the star had a large convective core during the first $\SI{\sim7.85}{\giga\year}$ of its evolution. Sub-solar mass stars like \kepl are born fully convective, but quickly retracts to only having a convective core when they reach the zero-age main-sequence. From standard evolution theory we expect the core to become radiative on short evolutionary time scales due to insufficient energy production, as energy is produced via the PP-chain, whereas super-solar mass stars produce energy via the more efficient CNO-cycle \citep{kw,kww}. The long-lived convective core from \buld therefore seems to contradict this. The long lifetime of the convective core in their model was produced by including convective overshoot which feeds the existing convective core material from the surrounding layers. This causes an increased energy production and thus requires energy to be transferred via convection for a longer part of its lifetime. Their analysis is elaborated further upon in \cref{subsec:buldgen_model}. 

This provides a unique problem to be studied. The question is not whether \kepl currently has a convective core as this is easily extracted from the frequency ratios to not be the case \citep{SilvaAguirre2011}. It is instead a question of whether the frequency ratios provide constraints on the lifetime of the initial convective core when it has long since then contracted and turned radiative. As the frequency ratios are a product of the current internal structure of a star, they will only contain signatures of a disappeared convective core if it is reflected in the internal structure. This signature would be caused by the difference in speed of which the convective core will retract, comparing to a short lifetime convective core model. This causes the composition gradient in the core to be essentially erased for an extended amount of time, thus affecting the sound speed gradient from which the frequency ratios are derived. While this can easily be modelled, the question remains whether this signature is strong enough to alter the frequency ratios significantly.

This work will focus on studying the survival of the birth convective core. Mainly, the lifetime of the convective core will be inferred from fitting the observed parameters of \kepl to a grid of stellar evolutionary tracks with high resolution sampling of convective core extent and lifetimes. The stellar models that best reproduce the observed constraints will then be compared to the results of \buld to investigate the origin of the inferred convective-core lifetime. Additionally, the fit of frequency ratios to the models of \buld will be reanalysed to determine which signatures/contributions lead to their conclusion of a long lifetime convective core being the best fit.

The paper is structured as follows. The observations of \kepl and previous modelling of the star serving as a basis for this paper is presented in \cref{sec:kepl}. The method for modelling the star is outlined in \cref{sec:modelling}, while the results from fitting the observations to these models are presented in \cref{sec:results}. The results will be discussed and compared to those of \buld in \cref{sec:discussion}, and the conclusion will be presented in \cref{sec:conclusion}.

\section{Previous studies}
\label{sec:kepl}

The following briefly outlines the observed parameters of \kepl and the analysis performed in \buld. 

\subsection{Observations}
\label{subsec:observations}

The first asteroseismic study of \kepl is that of \cite{Campante2015}, who determined individual oscillation modes and parameters from the \kepler light curve. The large frequency separation was determined to be $\dnu = \SI{179.64\pm 0.76}{\micro\hertz}$ and the frequency of maximum power to be $\numax = \SI{4538\pm 144}{\micro\hertz}$. Additionally, they used spectroscopic measurements from the Keck I telescope and HIRES spectrograph \citep{HIRES} to determine the effective temperature \teff, surface gravity \logg, and elemental abundances. Using these in the stellar modelling, they determined the star to be an old star with sub-solar mass.

\cite{Mack2018} has since performed a spectroscopic study of the star using higher resolution spectra obtained with the Potsdam Echelle Polarimetric and Spectroscopic Instrument \citep[PEPSI,][]{PEPSI}. From this they determined the effective temperature $T_{\textup{eff}}=\SI{5172\pm 75}{\kelvin}$, the metallicity $\left[\textup{Fe}/\textup{H}\right] = \SI{-0.52 \pm 0.12}{\dex}$, and the alpha-enhancement $\left[\alpha/\textup{Fe}\right] = 0.23 $ which are all used in this paper.

The asteroseismic observations of \kepl used in this work are those provided in \cite{Davies2016}. These include the frequencies of the individual oscillation modes and a quality estimation of these. While all determined frequency modes are reported, the quality assurance check flagged the two modes $\nu(\ell=2,n=24,25) = (\SI{4752.43}{\micro\hertz}, \SI{4932.4}{\micro\hertz})$ to be more likely to be noise in the data than detected frequency modes (i.e. prior dominated). These are referred to as the flagged modes (FM) in the following analysis where their inclusion will be varied. While they are flagged due to the quality estimation, it is also important to notice that they form the two highest $r_{02}$ frequency ratios, who are susceptible to shifts due to magnetic activity, as detailed by \cite{Thomas2021}. They may therefore also be unreliable due to this effect, so varying their inclusion is also relevant due to this.

The frequencies are corrected for the Doppler shift arising from the line-of-sight velocity of the star $V_r$ in accordance with \cite{Davies2014}, who derived the correction to be a factor of $\left(1+V_r/c \right)$ with $c$ being the speed of light. The frequency ratios are calculated according to \cref{eq:ratios,eq:five_point,eq:d02} using \num{10000} Monte Carlo realisations to determine their correlations \citep[see section 4.1.3 of][]{BASTA,Verma2022}.

With the recent data release from the Gaia mission \citep[DR3,][]{GaiaMission,GaiaDR3}, the line-of-sight velocity of \kepl is measured to be $V_r = \SI{-120.784\pm0.255}{\kilo\meter\per\second}$. This also provides high-accuracy measurements of the parallax $\varpi = \SI{27.358\pm0.012}{\milli\arcsecond}$, the corresponding zero-point offset of \SI{-20.468}{\micro\arcsecond} \citep[derived using the \texttt{gaiadr3-zeropoint} package,][]{Lindegren2021}, along with its coordinates. The observed magnitudes that will be used in conjunction with these are that of the Tycho-2 catalogue \citep{Tycho21,Tycho22}, which provides the apparent magnitudes $B_T = \num{9.898\pm0.024}$ and $V_T = \num{8.925\pm0.015}$.

Across the previous asteroseismic studies of \kepl \citep{Campante2015,SilvaAguirre2015,Buldgen2019}, they all determine fundamental stellar properties that agree within their uncertainties. It is determined that it has an age of \SI{\sim11\pm 1}{\giga\year}, has a mass of \SI{\sim0.748\pm0.044}{\solarmass} and a radius of \SI{\sim0.75\pm0.01}{\solarradius}.

\subsection{Forward modelling from \buld}
\label{subsec:buldgen_model}

\begin{figure}
    \centering
    \includegraphics[width=\linewidth]{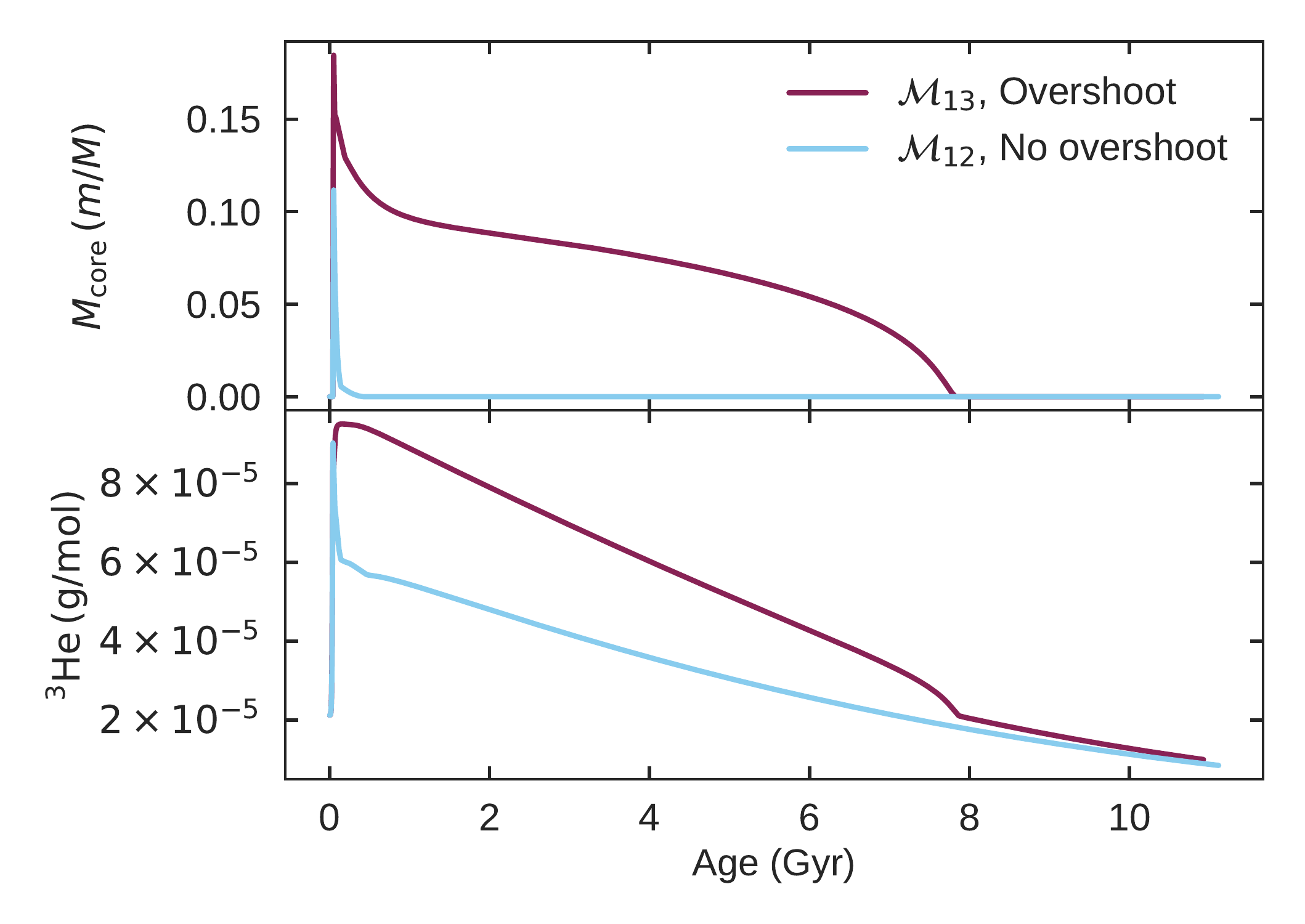}
    \caption{The evolution of \bm{12} (blue) and \bm{13} (red) in terms of the convective core size (\textit{top}) and \ce{^3He} abundance (\textit{bottom}). For \bm{13} the convective core is clearly quite extensive until \SI{8}{\giga\year} coinciding with its surplus of \ce{^3He}, while \bm{12} follows the predicted evolution with the helium content approximately equal to the equilibrium amount for the full evolution (\cref{app:he3}). \textit{Source:} Reproduction of \protect\cite[fig. 7]{Buldgen2019}, data from G. Buldgen, private communication.}
    \label{fig:buld_profile}
\end{figure}

The modelling in \buld was performed in three steps (G. Buldgen, private communication). Firstly, they used the AIMS software \citep{AIMS} to match the observations to a pre-calculated grid of models using a single set of input physics. Secondly, using the best-fit and neighbouring models from this, they used inversions of stellar structure to determine a precise measurement of the mean density of \kepl. Finally, with this constraint on the mean density together with the observed parameters described in the following paragraph, local minimisation was then performed with varying input physics, in order to obtain a better best fit model. The minimisation was executed using a Levenberg-Marquardt algorithm \citep{Levenberg1944,Jake1988}, while the models were computed on-the-fly using the Liège stellar evolution code \citep[CLES,][]{CLES} and the Liège stellar oscillation code \citep[LOSC,][]{LOSC}.

They determined the observed frequency ratios from the individual frequencies from \cite{Campante2015}. While the same modes were available as those in \cite{Davies2016}, quality estimations of the frequencies were not provided, and \buld did therefore not exclude the two flagged modes $\nu(\ell=2,n=24,25)$. They performed minimisation of the $\chi^2$ error function in regards to the $r_{02}$ frequency ratios, \feh, \teff, $\log g$, an estimated mean density $\rho=\SI{2.495\pm0.05}{\gram\per\cubic\centi\meter}$, and a self-derived value of the photometric luminosity $\lphot=\SI{0.40\pm0.04}{\solarlum}$ from a combination of Gaia DR2 and 2MASS observations. This was done across varying choices of input physics (equation of state, opacity tables, etc.) used for the models, which produced 13 different combinations of results \citep[see tab.~1 of][]{Buldgen2019}. The determined parameters agreed within $1\sigma$ across all the combinations. However, of these 13 resulting models, the last two models, here referred to as \bm{12} and \bm{13}, are of interest in this work as they had the lowest overall $\chi^2$ values. These models have been obtained from G. Buldgen (private communication) for the analysis and comparisons presented in this work. The determined properties of \bm{12} were a mass of \SI{0.753\pm0.018}{\solarmass}, an age of \SI{11.13\pm0.57}{\giga\year}, an initial hydrogen mass fraction $X_{\rm ini}=\num{0.746}$, and an initial metal mass fraction $Z_{\rm ini}=\num{0.0076}$. Similarly, the mass, age, $X_{\rm ini}$ and $Z_{\rm ini}$ for \bm{13} were determined to be \SI{0.755\pm0.020}{\solarmass}, \SI{10.95\pm0.61}{\giga\year}, \num{0.750} and \num{0.0069} respectively.

It was determined that \bm{13} had a lower $\chi^2$ value than \bm{12}, and it is worth noting that this difference in $\chi^2$ is mainly attributed to the comparison of the frequency ratios. The main difference in input physics between these models were that the treatment of convective overshoot was included in \bm{13}, while it was not included for \bm{12}. This resulted in the convective core mass fraction evolving quite differently for these two models as shown in \cref{fig:buld_profile}. While they are both initially fully convective, \bm{13} has a substantial convective core that survives until \SI{\sim 7.85}{\giga\year}, while the convective core of \bm{12} completely disappears after the first brief fully convective phase. This therefore appears to be an unconventional result as theory predicts that \bm{12} would be the best model as sub-solar mass stars generally do not have convective cores during main-sequence evolution. As the result was based on the better fit of frequency ratios, and these are constructed to be sensitive to the interior structure of the star, it was therefore concluded that \bm{13} provides the best prescription for the evolution of \kepl's convective core. It is this result which will be studied in this work. 

The main difference between the stellar evolution code used in \buld and the code used in this work (see \cref{subsec:fg_theory}) lies in the mathematical prescription of convective overshoot. \buld used a step overshooting prescription that extends the boundary of the convective core as determined by the Schwarzschild criterion \citep{kww} by an amount depending on the local pressure scale height and a free (hyper)parameter $\alpha_\mathrm{ov}$ \citep[sec. 3.5]{CLES}. This extended region is thereby treated as a fully convective zone with instantaneous mixing. On the other hand, in this work we use exponential overshooting and treats mixing in all convective zones as diffusive (for a comprehensive comparison of the methods see \cite{Pedersen2018}, section 2.1 and 2.2).

The presence of convective overshoot were shown by \buld to increase the abundance of \ce{^3He} compared to the equilibrium amount (see \cref{app:he3}), which were thus concluded to have caused an extended lifetime of the convective core. The correlation can clearly be seen in \cref{fig:buld_profile} (and \cref{fig:he3_equilibrium}) and is indeed the cause of the extended lifetime of the convective core in the models as was also shown in a previous study by \cite{Deheuvels2010}. However, while the mixing was modelled using convective overshoot, it was not known whether this or alternative phenomena such as rotation was the true cause that produced the additional mixing. It is here worth noting that an extended convective-core lifetime in the models of \buld might also be partly induced from assuming instantaneous mixing of the convective zones. As mentioned in \cite{Zhang2022}, this would require extreme convective velocities in the core, in order to fully mix before their nuclear consumption time. 
\section{Grid-based modelling}
\label{sec:modelling}

We use the so-called \emph{grid-based} modelling approach. Here a grid of stellar models spanning the suitable parameter space is computed and the likelihood of the models are evaluated given a set of observations. Due to the unsolved degeneracies in the stellar models \citep[for an overview, see][]{basuchaplin2017}, multiple configurations of stellar properties can yield high likelihoods when compared to the observed data. This grid-based method allows us to better get an overview of the different islands of probability or overall different solutions of sets of stellar properties that fit the observations. This method also allows for fitting different sets of observations without the cost of recomputing new models, which can be especially costly for the high-dimensional parameter space used in this work (see \cref{subsec:free_parameters}). An in-house code package\footnote{Planned for eventual public release accompanying the \basta-code: \\ \url{https://github.com/BASTAcode}} is used to construct these grids, which utilises the Sobol quasi-random, low-discrepancy sequences to uniformly populate the parameter space \citep{Sobol1,Sobol2,Sobol6,Sobol4,Sobol5,Sobol3}.

The stellar evolution models are computed using \textsc{garstec} \citep{GARSTEC}. The code utilises a combination of the equation of state by the \textsc{opal} group \citep{OPAL_EOS_1,OPAL_EOS_2}, and the Mihalas-Hummer-Däppen equation of state \citep{MHD_EOS_1,MHD_EOS_2,MHD_EOS_3,MHD_EOS_4}. Atomic diffusion of elements are treated following the prescription by \cite{Thoul1994}. For the opacity, the compilations used are the high temperature opacities from the \textsc{opal} group \citep{OPAL_opacity_1,OPAL_opacity_2} and the low temperature opacities from \cite{Ferguson2005}. The \textsc{nacre} \citep{NACRE} cross sections are used for the nuclear energy generation rate, except for the \ce{^{14}N(p,\gamma)^{15}O} and \ce{^{12}C(\alpha,\gamma)^{16}O} reactions that are instead from \cite{Formicola2004} and \cite{Hammer2005} respectively. The theoretical oscillation frequencies of the models are computed using the Aarhus adiabatic oscillation package \citep[ADIPLS,][]{ADIPLS}. For the computation of synthetic magnitudes the bolometric corrections of \cite{Hidalgo2018} are used in order to fit the parallax.

Due to \kepl being $\alpha$-enhanced, the element abundances have to be corrected accordingly. The stellar mixtures used are calculated based on the solar mixture from \cite{Asplund2009}, by uniformly varying the relative abundance of the $\alpha$-elements in steps of \SI{0.1}{\dex} in $\left[\alpha/{\rm Fe}\right]$ \citep[see][section 3.2]{BASTA}. The opacity tables are also updated taking the new abundances into account.

\subsection{Convective overshooting and the geometric cut-off}
\label{subsec:fg_theory}
\begin{figure}
    \centering
    \includegraphics[width=\linewidth]{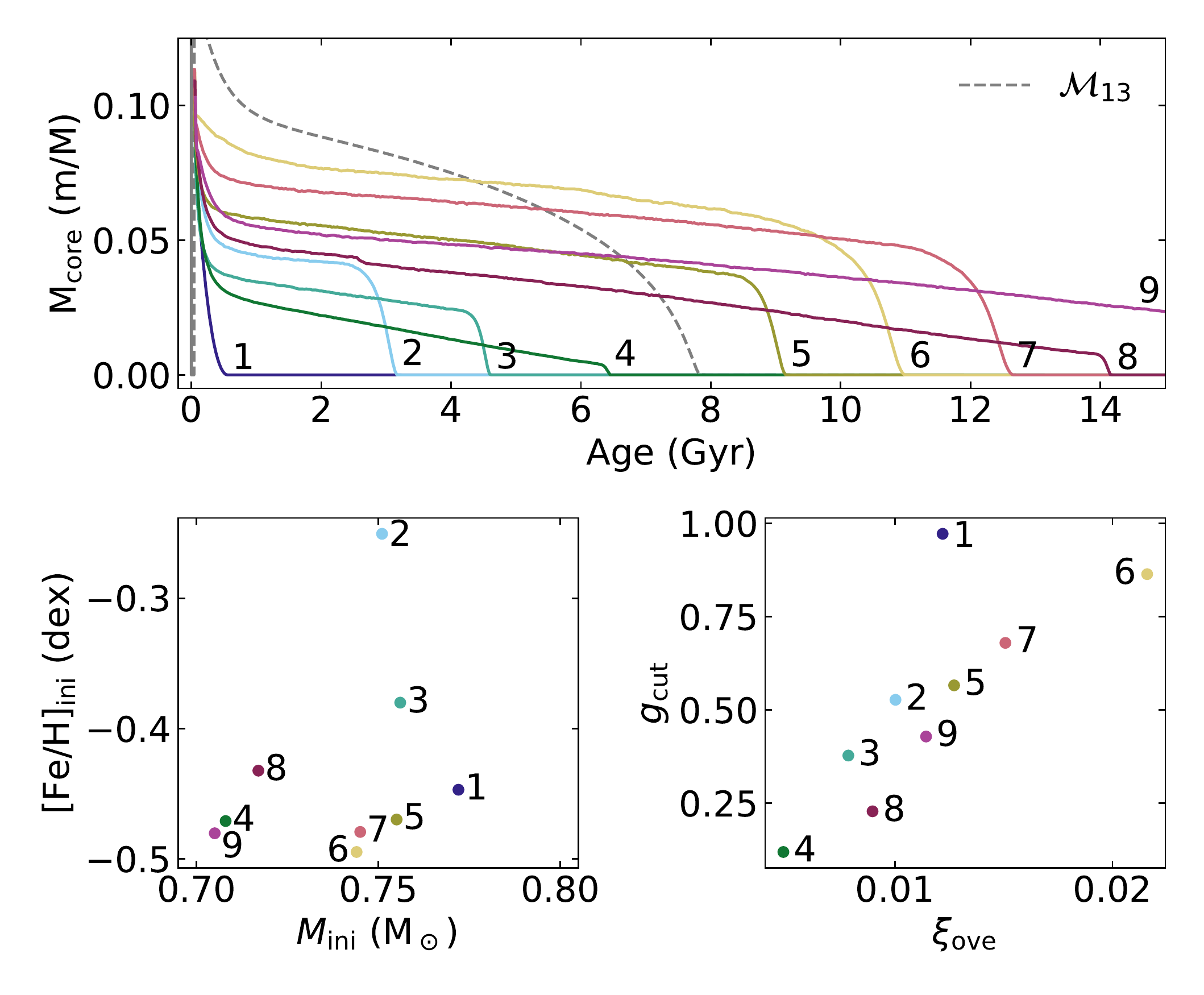}
    \caption{\textit{Top:} Fractional mass of the convective cores as a function of age for stellar evolution tracks with varied combinations of overshooting efficiency $f_{\rm ove}$ and cut-off coefficients $\gc$. The numbers label the tracks at the point their convective core disappears (\tcore) within the plotting range. \textit{Bottom left:} Initial mass and metallicity of the tracks. \textit{Bottom right:} Overshooting efficiency and cut-off coefficient of the tracks.}
    \label{fig:fg_vary}
\end{figure}
As mentioned previously, several different implementations of convective overshooting exist. The implementation used in this work is the default option in the stellar evolution code \textsc{garstec}. It treats overshooting as a diffusive process with the diffusion constant $D$ at a distance $z$ outside the convective zone being
\begin{align}
	D\left(z\right) = D_0 \exp \left(-\frac{2z}{f_{\rm ove}\tilde{H}_P}\right) \,,
	\label{eq:ove}
\end{align}
where $\tilde{H}_P$ is the so-called modified pressure scale height (see below), $D_0$ is the diffusive constant taken just inside the convective zone, and $f_{\rm ove}$ is the convective overshoot efficiency parameter, which is a dimensionless, free parameter of the model. 

The pressure scale height $H_P$ is locally modified to adjust the convective overshoot in correspondence with the size of the convective zone. It is scaled by the so-called geometric factor $G$, which is defined as
\begin{align}
    G = \left(\frac{R_\textup{cz}f_{\rm ove}}{\gc H_P}\right)^2 \,,
    \label{eq:geometric_factor}
\end{align}
where $R_\textup{cz}$ is the radial thickness of the convective zone, and \gc is the dimensionless cut-off coefficient. By default in \textsc{garstec}, the cut-off coefficient is set to produce the expected behaviour for the evolution of convective core in the Sun, which initialises it to $\gc=2$, but will here be treated as a free parameter in order to artificially induce longer convective-core lifetimes. To only reduce the amount of overshoot and not amplify it, $\tilde{H}_P$ is determined as
\begin{align}
	\tilde{H}_P = \begin{cases}
	H_P & \textup{if}\quad G\geq1 \\
	GH_P & \textup{if}\quad G<1\\
	\end{cases}\,.
	\label{eq:cutoff}
\end{align}
For low-mass stars and $\gc=2$ this results in the expected behaviour where no convective core is present after the first \SI{\sim200}{\mega\year} of evolution. On the other hand, lower values can result in convective overshoot to not be limited ($G\geq 1$), and thus allows for the core to be fed additional \ce{^3He}. This increases energy production and thus extends the lifetime of the convective core (see \cref{app:he3}). It will however still contract, but at a slower pace.

When the core has contracted to the point where $G<1$, the extent of the convective overshoot will be reduced following \cref{eq:ove,eq:cutoff}. Due to this, the amount of material the core is fed will also reduce, in turn causing further contraction of the convective core, and a reduction of the extent of the convective overshoot. This is therefore a self-reinforcing effect, which means that when the geometric cut-off is triggered, it will cause an exponential decrease in convective core size. 

\cref{fig:fg_vary} demonstrates how the convective core evolution depends on $f_{\rm ove}$ and \gc. Their evolution is summarised in terms of the convective-core lifetime \tcore, defined as the age where the convective core mass fraction first decreases below $10^{-4}$. Should the track end before this point, the core lifetime is estimated from a linear extrapolation to the last 50 models in the track, up to a maximum of \SI{20}{\giga\year}. The correlations between $f_{\rm ove}$ and \gc in \cref{fig:fg_vary} (bottom right panel), in relation to the convective core, can be described as follows:
\begin{itemize}
    \item The center of the plot shows models with a birth early convective core that disappears before its current age, and is thus the region where core evolution can be similar to that of \bm{13},
    \item low values of $f_{\rm ove}$ never produce convective cores of any significant size,
    \item large values of $f_{\rm ove}$ but high values of \gc produce a significant convective core that disappears rapidly ($<\SI{1}{\giga\year}$),
    \item large values of $f_{\rm ove}$ but low values of \gc produce a significant convective core that does not disappear before the current stellar age.
\end{itemize}

While the whole parameter space shown in \cref{fig:fg_vary} (bottom right panel) is sampled in the final grid, the tracks shown in the figure were selected to show a step-variation of \tcore, and it is therefore a biased selection. With enough variations of these hyper parameters, it should be possible to infer whether or not \kepl has had a convective core during its lifetime.

\subsection{Free parameters}
\label{subsec:free_parameters}

In this work, we use 7 free (initial) parameters for constructing the grid of models. They consist of the stellar mass, the parameters describing the composition of the star, and the (hyper) parameters of the hydrodynamics.

The composition of stars are determined from the initial metallicity $\feh_{\textup{ini}}$, the enhancement of alpha element abundance $[\alpha/\textup{Fe}]$, and the initial helium mass fraction $Y_{\textup{ini}}$. The initial hydrogen abundance $X_{\textup{ini}}$ and metal abundance $Z_{\textup{ini}}$ is derived from these and the solar metal to hydrogen mass fraction $(Z/X)_\odot$ of the given abundance table.

In one-dimensional models, the treatment of hydrodynamics is prescribed through the approximate mixing-length theory \citep[MLT, ][]{MLT}. It depends on the mixing-length parameter $\alpha_{\textup{MLT}}$, which describes the length $\ell$ over which a convective element inside the Schwarzschild border moves,
\begin{align}
    \ell = \alpha_{\textup{MLT}} H_P\,.
\end{align}
As it is not possible to derive the value of this parameter from first principles, it can either be calibrated to the solar value and assumed to be constant for all stars, or be allowed to vary as a free parameter. In this work, the latter approach is chosen. Additionally, the hyperparameters prescribing the convective overshoot, the already mentioned overshooting efficiency $f_{\rm ove}$ and the cut-off coefficient $\gc$, are also varied.

In total, these 7 free parameters will be varied when constructing the grid (as seen in \cref{tab:keplgrid} and the following subsection). The number of free parameters is considered when deciding upon the number of tracks in the grid in order to ensure a well distributed sampling along each dimension. 

\subsection{The grid of stellar models}
\label{subsec:kepl_grid}
\begin{table}
	\centering
	\caption{Varied parameters of the grid with \num{3000} tracks. The sampling is demonstrated in \cref{app:grid}.}
    \begin{threeparttable}
	\begin{tabular}{p{1.5cm} S[table-format = 2.2, table-column-width=1.5cm] S[table-format = 2.2, table-column-width=1.5cm]}
		\toprule
		Variable	&	{Minimum}	&	{Maximum}\\
		\midrule
		$M\, (\si{\solarmass})$	                &  0.70	&  0.80	\\
		$\var{[\textup{Fe}/\textup{H}]}{ini}$	& -0.55	& -0.25	\\
		$\var{Y}{ini}$		                    &  0.24	&  0.27	\\
		$[\alpha/\textup{Fe}]$				    &  0.2	&  0.3  \tnote{(a)}\\
		$\var{\alpha}{MLT}$	                    &  1.60	&  2.00	\\
		$f_{\rm ove}$		                    &  0.00	&  0.03	\\
		$\gc$                                   &  0.1  &  1.0  \\
		\bottomrule
	\end{tabular}
    \begin{tablenotes}
    \item[(a)] Opacity tables have only been computed in steps of 0.1, so only the two values are available in the grid.
    \end{tablenotes}
    \end{threeparttable}
	\label{tab:keplgrid} 
\end{table}

The grid of stellar models used to infer stellar properties of \kepl consists of 3000 stellar evolution tracks. The intervals of the parameters describing the tracks are given in \cref{tab:keplgrid}, and their variation is determined as described in the beginning of this section. Due to the multidimensional quasi-random designation of parameters, the best way to evaluate the occupancy of the parameter space is to determine the volume $w$ each point occupies in this space, which is then used to compute the Bayesian posterior (see \cref{subsec:BASTA}). The occupancy is also shown in \cref{app:grid}.

The range of models in a given track is limited between the zero-age main-sequence and when the model reaches a large frequency separation of $\Delta\nu = \SI{170}{\micro\hertz}$. Here, the zero-age main-sequence is determined as where the ratio between the hydrogen burning luminosity and the total luminosity reaches $1$. The maximum time step is between $20$ and \SI{40}{\mega\year}, depending on initial mass. Synthetic magnitudes are computed for the Tycho-2 photometric system and used for fitting the parallax, $\varpi$ \citep[for more details, see section 4.2.2 in][]{BASTA}. The grid has been computed with atomic diffusion enabled. 

Across the subset of parameter space spanned by $f_{\rm ove}$ and $\gc$, the size of the convective core can have varied initial sizes and evolve in a lot of different ways as demonstrated in \cref{fig:fg_vary}. Thus the grid has models available for all possible sizes and lifetimes of the early convective core, with a minimum $\min(\tcore)=\SI{0.139}{\giga\year}$ and a maximum, as described in \cref{subsec:fg_theory}, of $\max(\tcore) = \SI{20}{\giga\year}$. The grid contains 400 tracks with a convective-core lifetime $\tcore=[\num{1},\num{9}]\si{\giga\year}$.

\subsection{Fitting algorithm}
\label{subsec:BASTA}

To match the observations to the models, we use the BAyesian STellar Algorithm \citep[\basta,][\footnote{Available at: \url{https://github.com/BASTAcode/BASTA}}]{SilvaAguirre2015,BASTA}. It uses Bayes' theorem to combine prior knowledge about the stellar parameters $\bmath\Theta$ with the data $\bmath D$ to give the likelihood $P({\bmath D} | {\bmath \Theta})$ of observing the data given the model parameters. The total likelihood is the product of the likelihood of individual groups of observables, $\bmath{D}_i$, determined as
\begin{equation}
P({\bmath D}_i | {\bmath \Theta}) = \frac{1}{\sqrt{2\pi|\mathbfss{C}_i|}} \exp\left(-\chi_i^2 / 2\right),    
\label{eq:grouplike}
\end{equation}
where $|\mathbfss{C}_i|$ is the determinant of the covariance matrix, and 
\begin{equation}
\chi_i^2 = \frac{1}{N_i} \left(\bmath{O}_{i,{\rm obs}} - \bmath{O}_{i,{\rm mod}}\right)^{\rm T} \mathbfss{C}_i^{-1} \left(\bmath{O}_{{i,\rm obs}} - \bmath{O}_{i,{\rm mod}}\right),
\label{eq:chi2}
\end{equation}
is the chosen error function, where $N_i$ is the number of observables in the group, $\bmath{O}_{i,{\rm obs}}$ is the observed values of parameters in the group, and $\bmath{O}_{i,{\rm mod}}$ likewise for the model. Also included in the posterior is the prior information, here only the initial mass function \citep{Salpeter1955}. To marginalise the posterior probability, the weight of the model is taken into account. This weight is the product of the volume $w$ it occupies in the parameter space used to construct the grid (see \cref{subsec:kepl_grid}), and a measure of how much volume a model occupies along a track. The latter is determined from the difference in ages of the tracks, and therefore represents a measure of evolutionary speed \citep{Joergensen2005} (for a more detailed description of these calculations, see \cite{BASTA}, section~2).

The resulting marginalised posterior has no predetermined form, but instead accurately depicts the probability distribution for each parameter. This also means that more than one solution can appear, and one therefore needs to look at the posterior distributions to get a complete picture of the solution. To represent the distribution, the inferred parameters are traditionally reported in terms of the median value, and the 16 and 84 percentiles of the distribution, which is known as the \SI{68}{\percent} Bayesian credible interval.

One can also look at the best-fit model to evaluate properties such as the fit to the observed ratios. Here it is necessary to make the distinction between the two ways we define the best-fit model. Based on the computation of likelihood as just described, the \textit{highest-likelihood model} (HLM) can be extracted as the best-fit model. The second is the \textit{lowest-$\chi^2$ model} (LCM) as determined by the error function \cref{eq:chi2}, which is independent of the grid.The LCM definition of the best-fit model were also used in \buld and will be used for comparison between these works (see \cref{sec:discussion}).

A prominent advantage of using \basta is its versatility. It can handle the multitude of different observables needed in this study, which allows for testing of each observation's impact on the convective-core lifetime, which we will do in the following.
\section{Results}
\label{sec:results}
Here our independent study of \kepl is presented, both in terms of the overall determination of the properties of the star, and a deeper dive into the convective-core lifetime. Throughout this section, the HLM definition of the best-fitting model is used.

\subsection{General fit}
\label{subsec:general_fit}
\begin{figure}
    \centering
    \includegraphics[width=\linewidth]{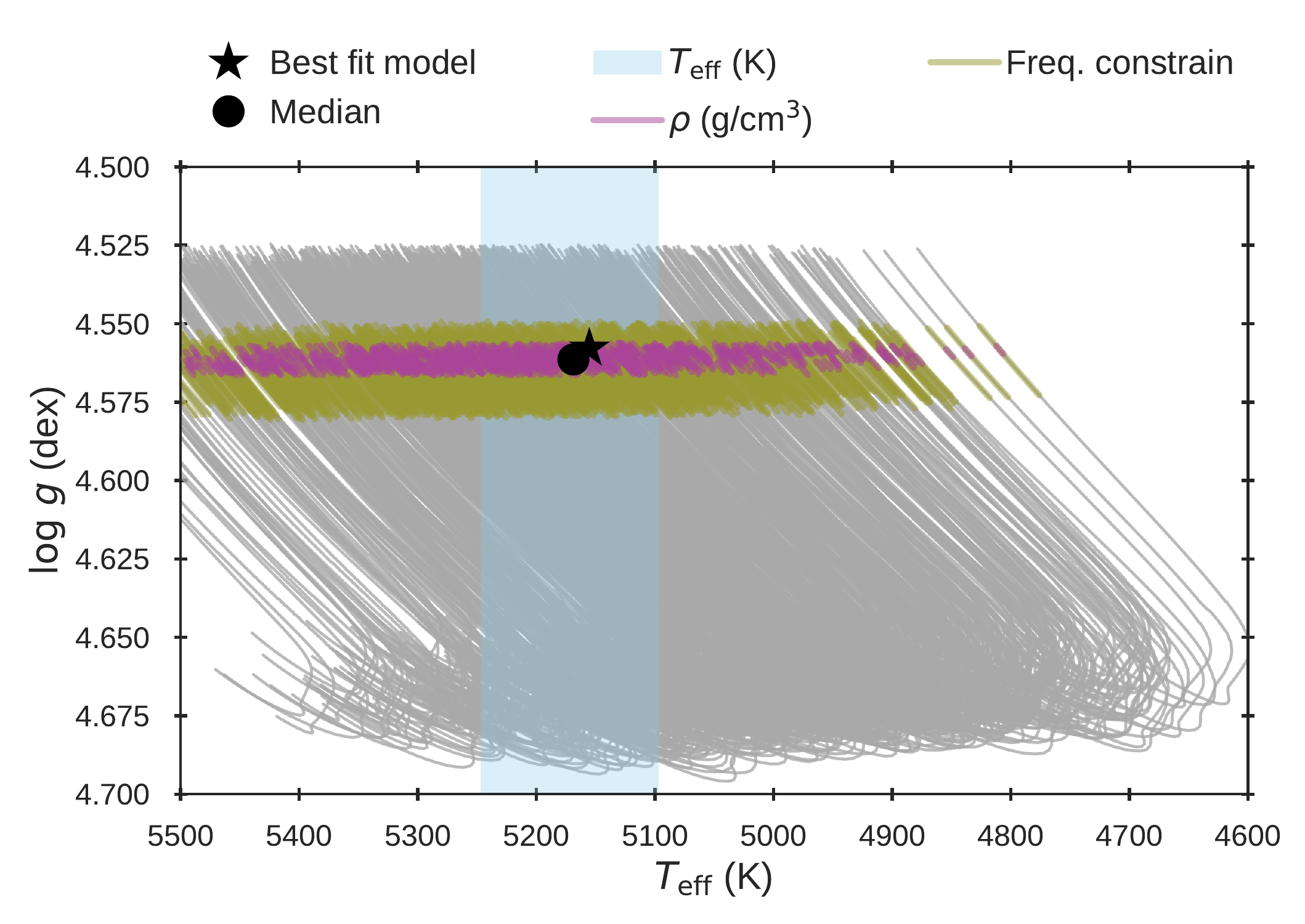}
    \caption{Kiel diagram of the fitted parameters across the evolutionary tracks in the grid. It shows the observational constraints of the fitted parameters within each track, as well as the median and best-fit model of the fit. The frequency constraint is described in the text, while the constraint from the observed metallicity is not shown as it barely changes along the track, but only tracks with models with a metallicity within the determined Bayesian credible interval is shown.}
    \label{fig:kiel}
\end{figure}
\begin{figure*}
    \centering
    \includegraphics[width=\linewidth]{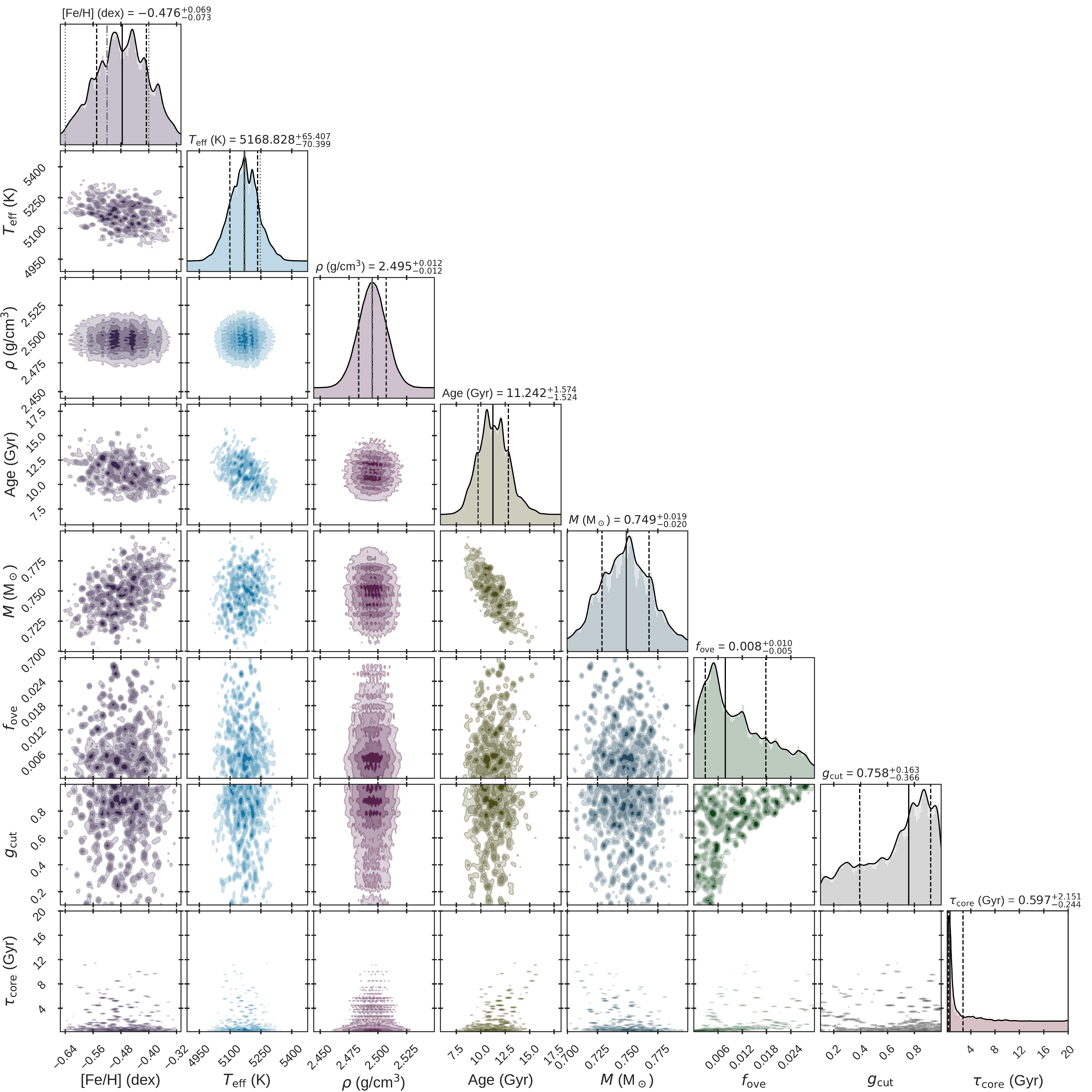}
    \caption{Corner diagram of the fitted and inferred parameters. The values are given (and shown in the diagonal panels) as the median (solid black line) and the 16th and 84th percentiles (dashed black lines) of the posteriors. On the diagonal, the grey lines gives the observed value (dash-dotted) with their uncertainties (dotted) for the fitted observables.}
    \label{fig:corner}
\end{figure*}

Here, we present the results of fitting the set of observables $\left(\feh, \teff, r_{012}, \rho\right)$. The usual chosen set of observables in \basta does not include $\rho$, as it is not readily observable, but is included here as an additional constraint on the interior structure derived from the individual frequencies. For testing and comparison, the full set as used in \buld as well as different sets of observables and definitions of the frequency ratio fit have been explored. These are presented in \cref{sec:discussion}. 

As described in \cref{subsec:observations}, the two frequency modes $\nu(\ell=2,n=24,25)$ have been flagged as unreliable. They have therefore been omitted in the following, whereby the two frequency ratios $r_{02}(\nu=\SI{4752.43}{\micro\hertz},\SI{4932.4}{\micro\hertz})$ are not included in the fit. These are referred to as the flagged modes (FM) in the following sections. The implications of this is explored further in \cref{sec:discussion}, as that were not the case for \buld.

The results are presented in \cref{fig:kiel,fig:corner}. \cref{fig:kiel} shows how the observed parameters overlap with the models in the grid by colouring the models with parameters within the observed limits. The frequency constraint in the figure is determined from matching the lowest $\nu(\ell=0)$ mode in the observations to the models in the grid. Only models having a frequency difference of this mode within $\pm 0.5\dnu$ are considered when calculating the likelihood, and it is this selection that is highlighted in \cref{fig:kiel}. Also shown is the median of the marginalised posterior shown in \cref{fig:corner}, and the best-fit model (here the HLM). \Cref{fig:kiel} shows that the observational constraints converge at a broad selection of models. While the constraints from the observed parameters are quite broad, compared to the resolution of the grid, the deviation of the best fit model from the median arises from the fit of ratios.

The marginalised posterior distributions of the fitted and inferred stellar properties are shown in the diagonal panels of \cref{fig:corner}, while 2-dimensional (2D) histograms of the likelihood between the parameters are shown in the remaining panels. The solution is in agreement with the previous studies of \kepl as presented in \cref{subsec:observations} and the fitted non-frequency parameters match well within their observed uncertainties. The posterior distributions mostly show unimodal distributions, so the properties are well constrained. It determines \tcore to be $0.597^{+2.151}_{-0.244}\,{\rm Gyr}$, with the 14th percentile being twice the minimum available \tcore of \SI{0.139}{\giga\year}.

The 2D-histogram between $f_{\rm ove}$ and \gc, (green histogram on the second-to-last row, sixth column) from the bottom left in \cref{fig:corner}, clearly shows no islands of probability in the lower-right region. Comparing this with the interpretation presented in \cref{subsec:fg_theory}, it clearly shows close-to-zero likelihood for any model that has a large convective core surviving until its current age. Meanwhile, the correlation does not directly describe the lifetime of a convective core that has disappeared before the current age. However, as can be seen from the \tcore histogram in \cref{fig:corner} (bottom right panel), the fit does infer a short-lived convective core. 

\subsection{Best-fit models}
\label{subsec:best_fit_models}
\begin{figure}
    \centering
    \includegraphics[width=\linewidth]{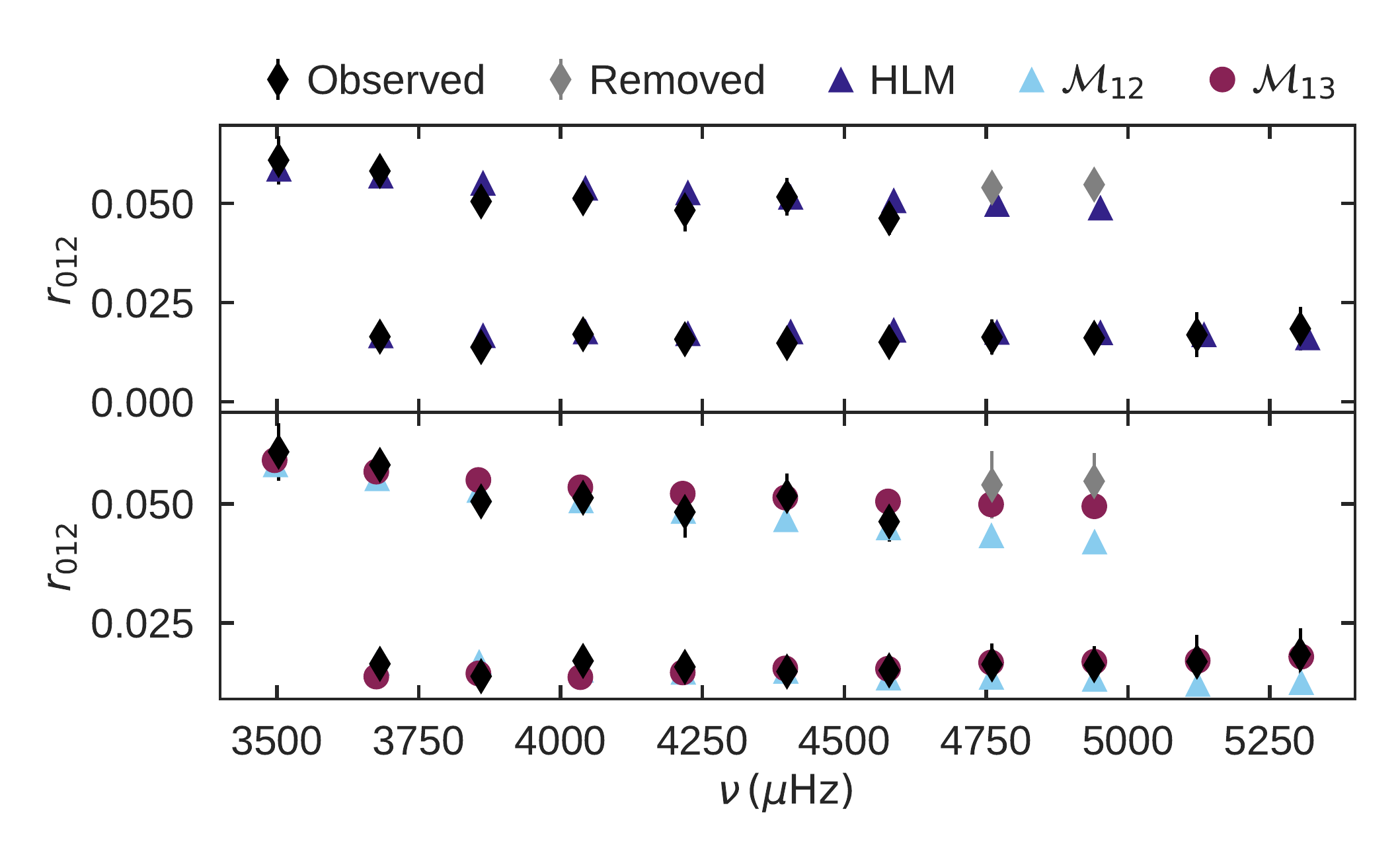}
    \caption{The $r_{012}$ frequency ratio sequence of the observed frequencies compared to the best-fit model using \basta (\textit{top}), and the \bm{12} and \bm{13} models from \buld (\textit{bottom}). The lower sequence in both panels is the $r_{01}$ ratios, while the upper sequence is the $r_{02}$ ratios, where the flagged modes $r_{02}\left(n=25,n=26\right)$ are shown in grey.}
    \label{fig:ratios}
\end{figure}

Shown in \cref{fig:ratios} is the ratios of the HLM compared to the observed ratios alongside the ratios of \bm{12} and \bm{13}. It can be seen that the flagged modes show a clear deviation from the $r_{02}$ sequence, being greater than the extrapolation of the sequence would predict. We here emphasise that the offset between the modes on the x-axis in the top panel is caused by only fitting the ratios and not the individual frequencies. The échelle diagrams of the HLM are shown in \cref{app:echelle}.

\begin{figure}
    \centering
    \includegraphics[width=\linewidth]{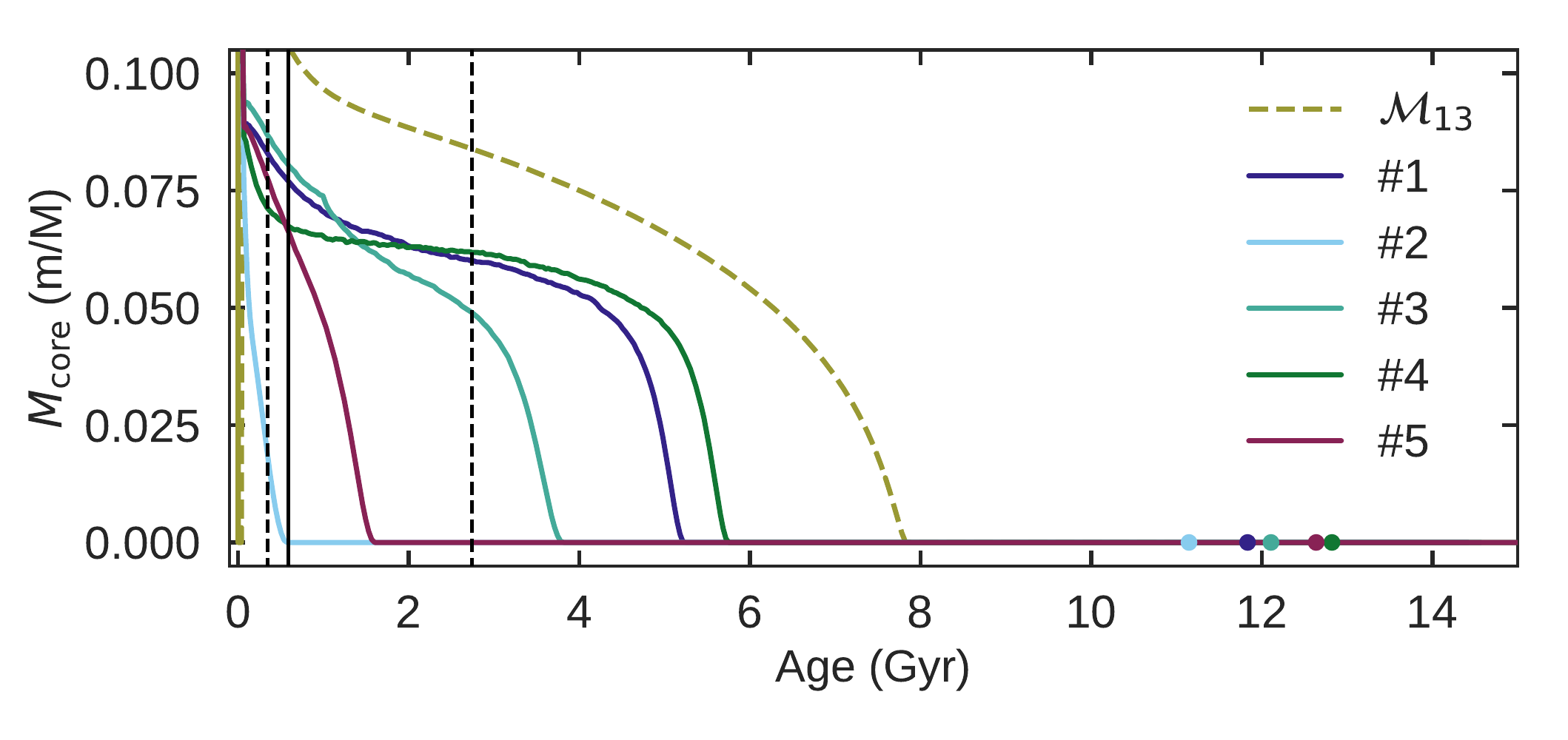}
    \caption{Convective core mass fraction as a function of age for the 5 best-fitting tracks determined from comparing the HLM within each track. The dots indicate the HLM within each track, while the black vertical lines show the inferred median (solid) and 16th and 84th percentile (dashed) \tcore from the posterior (bottom right panel of \cref{fig:corner}).}
    \label{fig:core_best}
\end{figure}

In order to get a better understanding of the inferred \tcore, we look into the convective core evolution of the best 5 tracks, defined by comparing the HLM within each track and here simply numbered using \# with $\#1$ being the best track using this metric. The convective core evolution of these 5 best tracks compared to \bm{13} is shown in \cref{fig:core_best}. It shows that it is not dominated by the expected interpretation of $\tcore<\SI{1}{\giga\year}$, but rather shows a spread around the upper limit. This indicates that the ratios do not provide a large enough constraint on the early convective core to differentiate between these values of \tcore. This might be due to how the prolonged \tcore is modelled or the precision of the observed modes, but it is more likely that \tcore is a parameter not generally well-constrained by the ratios, which is discussed further in \cref{subsec:fitvar}.

\begin{figure}
    \centering
    \includegraphics[width=\linewidth]{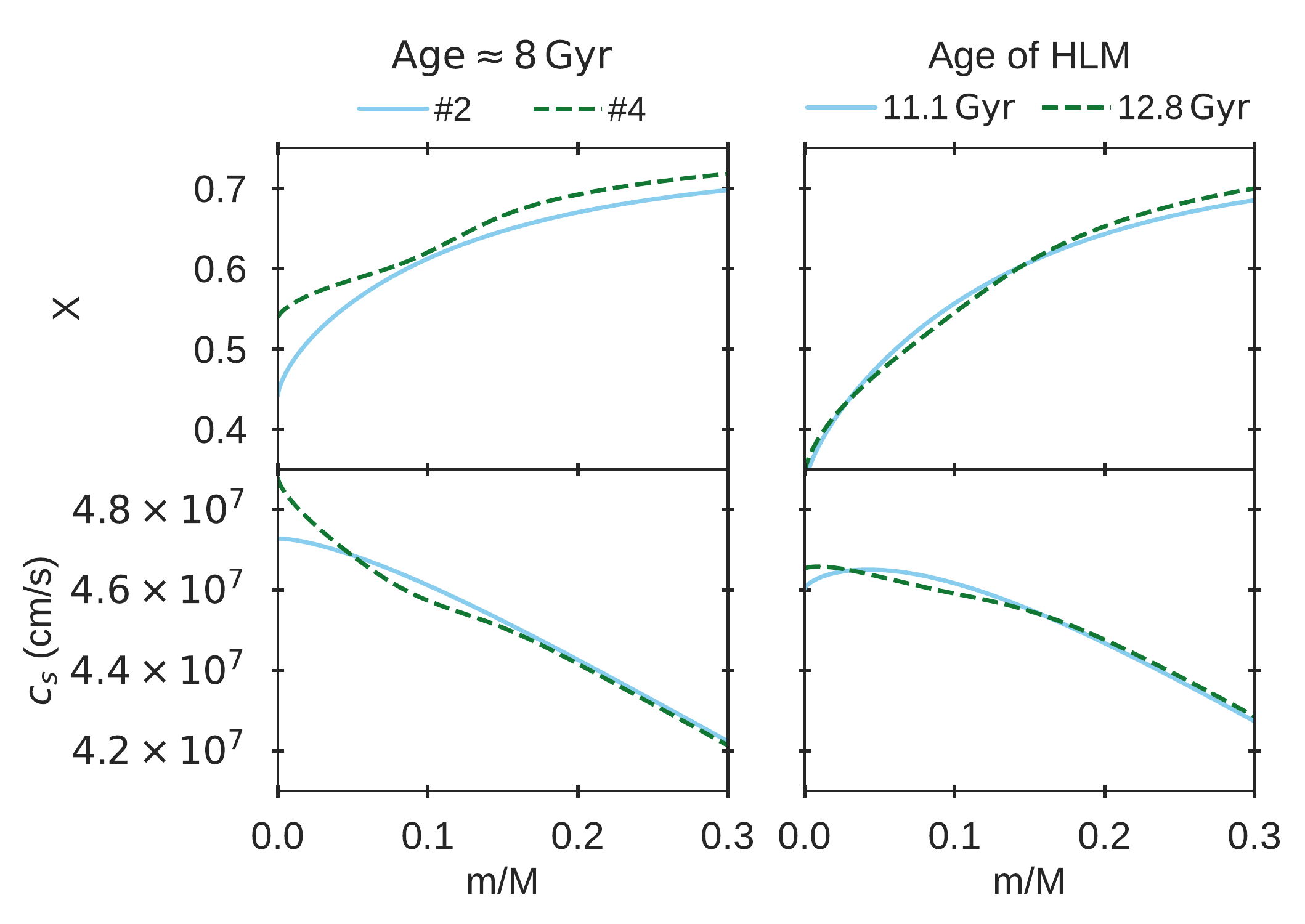}
    \caption{The local hydrogen mass fraction (\textit{top}) and sound speed (\textit{bottom}) profiles as a function of mass of the two tracks with the greatest difference in core lifetime from \cref{fig:core_best}. \textit{Left:} The profiles of models within the track at $\approx \SI{8}{\giga\year}$. \textit{Right:} The models within the track with the highest likelihood (HLM, highest likelihood model), at the age given in the legend.}
    \label{fig:profile_differences}
\end{figure}

To explore this spread in \tcore of the HLM's, the tracks with the largest difference in \tcore among these 5 best-fitting tracks in \cref{fig:core_best}, track $\#2$ with the lowest \tcore and track $\#4$ with the highest, are explored in more detail. As the modelled ratios are effectively a function of the sound speed $c_s$ in the stellar interior and thereby the profile in chemical composition of the model \citep{Aerts2010}, looking at these profiles should give some understanding of the issue in constraining \tcore. The difference between the local hydrogen mass fraction $X$ and sound speed profile of the two tracks are shown in \cref{fig:profile_differences} at an age of \SI{8}{\giga\year} and at the age of their respective highest likelihood model. While it is apparent at \SI{8}{\giga\year} that track $\#4$ still carries a signature of a convective core in both profiles, it is no longer apparent at the HLM, where it is similar to the short \tcore track. 

The highest likelihood model of track $\#4$ is however much older than that of $\#2$. This indicates that a reason why large \tcore models in the fit have an adequately high likelihood is that they have evolved for longer until the traces of the convective core has essentially disappeared from their interior structure. As the frequency ratios are the best parameter for constraining the age within the available observations, this allows for the inference of older models, compared to low \tcore models, which obtain a profile without traces of a convective core much earlier. It is also apparent in \cref{fig:core_best}, as the age of the HLM somewhat scales with \tcore. This is also vaguely apparent in the 2D histogram between age and \tcore in \cref{fig:corner} (bottom row, column 3), as a linear upwards trend, but is smeared out due to the degeneracy of the age with variation of the other parameters in the grid. 

\subsection{Additional models}
The presented analysis has also been repeated using other models of varied input. The general results are here briefly summarised.

An equally extensive grid of models as presented in \cref{subsec:kepl_grid} without the inclusion of atomic diffusion, and an appropriately adjusted initial metallicity, has been tested. These models do however not reliably match the observations, as is expected, as it is generally accepted that diffusion is needed to produce realistic models of sub-solar mass stars \citep[e.g. ][]{JCD1993,SilvaAguirre2017,Sahlholdt2018}.

To verify the approach of varying the geometric cut-off, as described in \cref{subsec:fg_theory}, two additional cases have been studied. The first case completely removes the geometric cut-off, which however allowed for convective cores to fluctuate in and out of existence throughout the whole evolution due to numerical effects. These were therefore not used for the analysis. 

Additionally, while the default value of \gc in the \textsc{garstec} code is 2, the grid used in this analysis were chosen to have a maximum value of 1. This was selected based on initial tests with varying \gc between 1 and 2 (along with a varying of $f_{\rm ove}$), which were determined to only produce models with $\tcore < \SI{1}{\giga\year}$. This interval would therefore be redundant to sample in the grid, and was therefore left out.
\section{Discussion}
\label{sec:discussion}
As the lifetime of the convective core we determined in \cref{sec:results} is different compared to that of \buld, it is of interest to determine what leads to this difference. Here we vary the choice of observables to use in the fitting set and the choices related to fitting the frequency ratios and compare for both the method presented in this work and for \bm{12} and \bm{13} from \buld to determine their influence on \tcore.

The comparisons are based on comparing $\chi^2$ values. As described in \cref{subsec:BASTA}, these are also determined for each model in this work in accordance with \cref{eq:chi2}, which can therefore be used to compare to the results of \buld. For consistency, the model values and frequencies of \bm{12} and \bm{13} are here used to re-compute $\chi^2$ values within the same framework. The mean density was, however, not available in the models, and have instead been estimated to be $\rho_{\bm{12}} = \SI{2.4925}{\gram\per\centi\meter\cubed}$ and $\rho_{\bm{13}} = \SI{2.4933}{\gram\per\centi\meter\cubed}$ from fig.~3 in \buld. 

The $\chi^2$ determined from \cref{eq:chi2} is only for a given set of observables, where the correlations are known. For this study, this only includes the ratios as the correlations of the other observables are not known. Therefore, for transparency, the total $\chi^2$ of a model is determined from the ratio contribution $\chi_r^2$ calculated from \cref{eq:chi2} along with the remaining observables $i$ as
\begin{align}
    \chi^2 = \chi_r^2 + \sum_i \left(\frac{O_{i,{\rm obs}} - O_{i,{\rm mod}}}{\sigma_i}\right)^2\,,
    \label{eq:chi2sum}
\end{align}
where $\sigma_i$ is their observational uncertainty. Note that this implies that the contribution to the total value from the ratios are divided by the number of ratios ($N_i$ in \cref{eq:chi2}), and are thus weighted equally with a single classical global parameter as is default in \basta. It is possible to weigh each ratio as an individual measurement \citep{PLATOH&H,Verma2022}, and the effect of this is detailed at the end of \cref{subsec:fitvar}. 

Finally, it is important to note here that \buld used the frequencies from \cite{Campante2015} to determine the frequency ratios, while we here only use the frequencies from \cite{Davies2016}. The $\chi^2$ values will therefore be close to, but not exactly the same values as in \buld.

In the following, the variation of $\chi_r^2$ will be discussed separately from the variation of the total $\chi^2$, which is discussed thereafter.

\subsection{Impact of ratio fitting}
\label{subsec:chi2r}
\begin{table*}
	\centering
	\caption{Our calculated $\chi_r^2$ values of \bm{12} and \bm{13} for the 16 different cases of fitting the ratios, where the lowest value is highlighted in bold font. The ``Seq.'' column denotes the frequency ratio sequence fitted. The $3p/5p$ column denotes whether the three- or five-point definitions of the small frequency separations (\cref{eq:three_point,eq:five_point} respectively) have been used, and it is left blank when fitting the $r_{02}$ sequence as it does not change depending on this definition. The FM (Flagged Modes) column denotes when the two ratios $r_{012}\left(n=25,26\right)$ have been excluded, while crossed out when they are included and left blank for the $r_{01}$ sequence as they are not used for this sequence. The ``Cov.'' column denotes when the covariance of the frequency ratios has been taken into account in the $\chi_r^2$ calculation and crossed out when the correlations are not taken into account. The last two columns lists the $\chi_r^2$ for the \norat case, see text for details.}
	\begin{tabular}{S[table-format=2.0] c c c c c c c c}
		\toprule
		{Case} & $\chi_r^2(\bm{12})$ & $\chi_r^2(\bm{13})$ & Seq. & $3p/5p$ & FM & Cov. & $\chi_r^2(\bm{12},\norat)$ & $\chi_r^2(\bm{13},\norat)$ \\
		\midrule
         1 & \textbf{1.111} & 1.518 & $r_{01}$ & $3p$ &  & \transparent{0.4}x  & 0.721 & \textbf{0.484}\\
         2 & \textbf{1.070} & 1.284 & $r_{01}$ & $5p$ &  & \transparent{0.4}x  & 0.743 & \textbf{0.170}\\
         3 & \textbf{1.083} & 1.426 & $r_{01}$ & $3p$ &  & \checkmark  & 0.666 & \textbf{0.364}\\
         4 & \textbf{1.446} & 1.908 & $r_{01}$ & $5p$ &  & \checkmark  & 0.686 & \textbf{0.219}\\
         5 & 1.215 & \textbf{0.942} & $r_{02}$ & & \transparent{0.4}x & \transparent{0.4}x  & 1.225 & \textbf{0.926}\\
         6 & \textbf{0.403} & 0.942 & $r_{02}$ & & \checkmark & \transparent{0.4}x  & \textbf{0.411} & 0.967\\
         7 & 1.234 & \textbf{0.922} & $r_{02}$ & & \transparent{0.4}x & \checkmark  & 1.198 & \textbf{0.907}\\
         8 & \textbf{0.404} & 0.959 & $r_{02}$ & & \checkmark & \checkmark  & \textbf{0.412} & 0.965\\
         9 & \textbf{1.161} & 1.284 & $r_{012}$ & $3p$ & \transparent{0.4}x & \transparent{0.4}x  & 0.940 & \textbf{0.670}\\
        10 & 1.141 & \textbf{1.110} & $r_{012}$ & $5p$ & \transparent{0.4}x & \transparent{0.4}x  & 0.965 & \textbf{0.529}\\
        11 & \textbf{0.843} & 1.323 & $r_{012}$ & $3p$ & \checkmark & \transparent{0.4}x  & \textbf{0.606} & 0.654\\
        12 & \textbf{0.801} & 1.136 & $r_{012}$ & $5p$ & \checkmark & \transparent{0.4}x  & 0.607 & \textbf{0.497}\\
        13 & \textbf{1.211} & 1.991 & $r_{012}$ & $3p$ & \transparent{0.4}x & \checkmark  & 0.902 & \textbf{0.665}\\
        14 & \textbf{1.477} & 2.386 & $r_{012}$ & $5p$ & \transparent{0.4}x & \checkmark  & 0.921 & \textbf{0.721}\\
        15 & \textbf{0.950} & 2.068 & $r_{012}$ & $3p$ & \checkmark & \checkmark  & \textbf{0.578} & 0.630\\
        16 & \textbf{1.168} & 2.515 & $r_{012}$ & $5p$ & \checkmark & \checkmark  & \textbf{0.579} & 0.661\\
		\bottomrule
	\end{tabular}
	\label{tab:chi2r}
\end{table*}
The better $\chi^2$ of the fit to the model with a convective core \bm{13} in \buld is attributed to the ratios. There are however several choices behind how the frequency ratios are fitted. Here, we explore these to determine how they affect the difference in $\chi_r^2$ between the models.

While the $r_{01}$ ratios in this work so far uses the five-point small frequency separations (\cref{eq:five_point}), one can also use the less correlated but also less smooth three-point small frequency separations (\cref{eq:three_point}). Both have been implemented and will be compared, referred to as the $5p$ and $3p$ cases respectively. Secondly, it is varied whether the two uncertain ratio modes $r_{02}\left(n=25,26\right)$ flagged by \cite{Davies2016} are included. It is apparent from \cref{fig:ratios} that these modes match \bm{13} better than \bm{12}, so their inclusion is varied in order to compare the influence on the fit. Lastly, it is varied whether the covariance $\mathbfss{C}_r$ of the determined ratios has been taken into account or not. Frequency ratios are by design correlated as they are computed from the same set of mode frequencies so this can have a large impact.

Our calculated $\chi^2_r$ values of \bm{12} and \bm{13} for all the combinations of the above mentioned choices are compared in \cref{tab:chi2r}, where case 5 corresponds to the choices made in \buld. As shown, the conclusion that the ratios favour \bm{13} over \bm{12} is only recovered for cases 5, 7, and 10, and can be attributed to two effects. Firstly, it is only recovered when the two flagged frequency modes are included in the set of frequencies, demonstrating the importance of well-determined mode frequencies, outside of their reported uncertainties. The other effect is the inclusion of covariance in the computation as it generally increases the $\chi^2$ value due to the ratios being highly correlated by construction, comparing e.g.\@ case 9 to 13, and 11 to 15. The increase in $\chi^2$ value is also larger for \bm{13} than \bm{12}. 

\begin{table*}
    \centering
    \caption{Comparison of $\chi^2$ values between the fits using \basta and the \bm{12} and \bm{13} models. HLM refers to the highest likelihood model in the \basta fit and LCM refers to the lowest $\chi^2$ model in the \basta fit as introduced in \cref{subsec:BASTA}. \tcore is the determined core lifetime from the fit. Fitting set describes the non-seismic fitted parameters, apart from \feh and \teff which are always included in the fit. Their values and uncertainties are detailed in \cref{subsec:observations}. The last four columns follow the same definitions as in \cref{tab:chi2r}. 
    }
    \begin{tabular}{S[table-format=2.0] c c c c c l c c c c c c}
        \toprule
        {Case} & {$\chi^2({\rm HLM})$} & {$\chi^2({\rm LCM})$} & {$\chi^2(\bm{12})$} & {$\chi^2(\bm{13})$} & {$\tau_{\rm core}\,[\si{\giga\year}]$}  & \multicolumn{1}{c}{Fitting set} &	Seq.	&   $3p/5p$ &   FM &   Cov. \\
        \midrule

         1 & 0.899 & 0.535 & 1.338 & 2.415 & ${0.631}_{-0.261}^{+2.700}$ &  & $r_{01}$ & $3p$ &  & \transparent{0.4}x \ts\\
         2 & 0.772 & 0.485 & 1.297 & 2.180 & ${0.641}_{-0.264}^{+2.773}$ &  & $r_{01}$ & $5p$ &  & \transparent{0.4}x \ts\\
         3 & 0.610 & 0.526 & 1.310 & 2.322 & ${0.631}_{-0.262}^{+2.778}$ &  & $r_{01}$ & $3p$ &  & \checkmark \ts\\
         4 & 1.014 & 0.614 & 1.672 & 2.804 & ${0.631}_{-0.263}^{+2.706}$ &  & $r_{01}$ & $5p$ &  & \checkmark \ts\\
         5 & 1.037 & 0.684 & 1.441 & 1.838 & ${0.767}_{-0.368}^{+6.366}$ &  & $r_{02}$ & & \transparent{0.4}x & \transparent{0.4}x \ts\\
         6 & 0.963 & 0.433 & 0.629 & 1.838 & ${0.713}_{-0.313}^{+5.307}$ &  & $r_{02}$ & & \checkmark & \transparent{0.4}x \ts\\
         7 & 1.029 & 0.681 & 1.461 & 1.819 & ${0.767}_{-0.368}^{+6.366}$ &  & $r_{02}$ & & \transparent{0.4}x & \checkmark \ts\\
         8 & 0.981 & 0.446 & 0.631 & 1.856 & ${0.713}_{-0.313}^{+5.307}$ &  & $r_{02}$ & & \checkmark & \checkmark \ts\\
         9 & 1.025 & 0.687 & 1.388 & 2.180 & ${0.634}_{-0.257}^{+2.779}$ &  & $r_{012}$ & $3p$ & \transparent{0.4}x & \transparent{0.4}x \ts\\
        10 & 0.980 & 0.677 & 1.367 & 2.006 & ${0.636}_{-0.256}^{+2.777}$ &  & $r_{012}$ & $5p$ & \transparent{0.4}x & \transparent{0.4}x \ts\\
        11 & 0.850 & 0.515 & 1.069 & 2.219 & ${0.635}_{-0.256}^{+2.465}$ &  & $r_{012}$ & $3p$ & \checkmark & \transparent{0.4}x \ts\\
        12 & 0.809 & 0.477 & 1.028 & 2.033 & ${0.634}_{-0.254}^{+2.466}$ &  & $r_{012}$ & $5p$ & \checkmark & \transparent{0.4}x \ts\\
        13 & 1.022 & 0.642 & 1.438 & 2.887 & ${0.624}_{-0.253}^{+2.438}$ &  & $r_{012}$ & $3p$ & \transparent{0.4}x & \checkmark \ts\\
        14 & 1.139 & 0.734 & 1.703 & 3.283 & ${0.613}_{-0.243}^{+2.329}$ &  & $r_{012}$ & $5p$ & \transparent{0.4}x & \checkmark \ts\\
        15 & 0.809 & 0.466 & 1.177 & 2.965 & ${0.622}_{-0.245}^{+2.195}$ &  & $r_{012}$ & $3p$ & \checkmark & \checkmark \ts\\
        16 & 0.894 & 0.527 & 1.395 & 3.412 & ${0.625}_{-0.248}^{+2.123}$ &  & $r_{012}$ & $5p$ & \checkmark & \checkmark \ts\\
        17 & 2.003 & 1.489 & 1.846 & 2.288 & ${0.631}_{-0.260}^{+2.548}$ & \lphot & $r_{012}$ & $3p$ & \transparent{0.4}x & \transparent{0.4}x \ts\\
        18 & 1.938 & 1.500 & 1.853 & 3.519 & ${0.609}_{-0.239}^{+2.074}$ & \lphot & $r_{012}$ & $5p$ & \checkmark & \checkmark \ts\\
        19 & 1.027 & 0.689 & 1.716 & 2.490 & ${0.634}_{-0.257}^{+2.779}$ & $\log g$ & $r_{012}$ & $3p$ & \transparent{0.4}x & \transparent{0.4}x \ts\\
        20 & 0.894 & 0.528 & 1.723 & 3.721 & ${0.624}_{-0.247}^{+2.124}$ & $\log g$ & $r_{012}$ & $5p$ & \checkmark & \checkmark \ts\\
        21 & 1.172 & 0.862 & 1.473 & 2.231 & ${0.597}_{-0.249}^{+2.734}$ & $\rho$ & $r_{012}$ & $3p$ & \transparent{0.4}x & \transparent{0.4}x \ts\\
        22 & 0.874 & 0.776 & 1.480 & 3.462 & ${0.592}_{-0.244}^{+2.150}$ & $\rho$ & $r_{012}$ & $5p$ & \checkmark & \checkmark \ts\\
        23 & 2.003 & 1.490 & 2.174 & 2.597 & ${0.631}_{-0.260}^{+2.548}$ & $\log g$,\,\lphot & $r_{012}$ & $3p$ & \transparent{0.4}x & \transparent{0.4}x \ts\\
        24 & 1.938 & 1.501 & 2.181 & 3.829 & ${0.610}_{-0.239}^{+2.074}$ & $\log g$,\,\lphot & $r_{012}$ & $5p$ & \checkmark & \checkmark \ts\\
        25 & 2.116 & 1.524 & 1.931 & 2.338 & ${0.579}_{-0.234}^{+2.169}$ & \lphot,\,$\rho$ & $r_{012}$ & $3p$ & \transparent{0.4}x & \transparent{0.4}x \ts\\
        26 & 2.036 & 1.532 & 1.938 & 3.570 & ${0.579}_{-0.232}^{+1.970}$ & \lphot,\,$\rho$ & $r_{012}$ & $5p$ & \checkmark & \checkmark \ts\\
        27 & 1.172 & 0.862 & 1.801 & 2.541 & ${0.597}_{-0.249}^{+2.734}$ & $\log g$,\,$\rho$ & $r_{012}$ & $3p$ & \transparent{0.4}x & \transparent{0.4}x \ts\\
        28 & 0.874 & 0.776 & 1.808 & 3.772 & ${0.592}_{-0.244}^{+2.150}$ & $\log g$,\,$\rho$ & $r_{012}$ & $5p$ & \checkmark & \checkmark \ts\\
        29 & 2.117 & 1.525 & 2.259 & 2.648 & ${0.579}_{-0.233}^{+2.169}$ & $\log g$,\,\lphot,\,$\rho$ & $r_{012}$ & $3p$ & \transparent{0.4}x & \transparent{0.4}x \ts\\
        30 & 2.036 & 1.533 & 2.266 & 3.879 & ${0.579}_{-0.232}^{+1.970}$ & $\log g$,\,\lphot,\,$\rho$ & $r_{012}$ & $5p$ & \checkmark & \checkmark \ts\\
        31 & 2.278 & 1.578 & 2.312 & 2.306 & ${0.677}_{-0.314}^{+5.342}$ & $\log g$,\,\lphot,\,$\rho$ & $r_{02}$ & & \transparent{0.4}x & \transparent{0.4}x \ts\\
        32 & 1.320 & 1.119 & 1.502 & 2.323 & ${0.631}_{-0.267}^{+3.962}$ & $\log g$,\,\lphot,\,$\rho$ & $r_{02}$ & & \checkmark & \checkmark \ts\\
        33 & 1.220 & 0.692 &  &  & ${0.633}_{-0.251}^{+3.119}$ & $\varpi$ & $r_{012}$ & $3p$ & \transparent{0.4}x & \transparent{0.4}x \ts\\
        34 & 1.122 & 0.531 &  &  & ${0.612}_{-0.231}^{+2.182}$ & $\varpi$ & $r_{012}$ & $5p$ & \checkmark & \checkmark \ts\\

        \bottomrule
    \end{tabular}
    \label{tab:chi2full}
\end{table*}
However, when looking at the $r_{01}$ sequence exclusively, it becomes obvious that the $\chi^2$ contribution of the mode $r_{01}(n=21)$ dominates the $\chi_r^2$ for \bm{13}, due to its small observational uncertainty, while the corresponding mode of \bm{13} also deviates relatively much. Here we explore how the results are affected by this mode. The fitting case \norat is introduced where this mode is excluded from the $\chi_r^2$ computation. The last two columns in \cref{tab:chi2r} shows the comparison between the $\chi^2_r$ of \bm{12} and \bm{13} for this case, for all of the combinations of fitting ratios described before. It shows that the $r_{01}$ sequence fits \bm{13} much better when excluding this mode, which is also reflected in some cases of fitting the full $r_{012}$ sequence. However, the conclusion for our preferred fitting case (case 16) remains unchanged, although the difference in $\chi^2$ between the models is significantly lower.

Therefore, the combined choices of inclusion of the flagged modes and not fitting the covariance in the ratio fitting appear to cause the preference of \bm{13} over \bm{12}. In our preferred choice of the ratio fitting framework, case 16, we therefore find \bm{12} to be a better fit to the observations than \bm{13} based on the frequency ratios alone. 

\subsection{Choice of fitting parameters}
\label{subsec:fitvar}

To make a realistic comparison to the work of \buld, and determine the influence on the inferred \tcore, we repeat the fit presented in \cref{sec:results} using different combinations of fitting sets and ratio fitting choices. 

First of all, to test whether a sufficient density of the grid has been reached, the fit from \cref{sec:results} has been repeated using only the first half (1500 tracks) of the grid. This fit did not determine a result noticeably different from what was obtained using the whole grid. We can therefore conclude that the grid has reached a sufficient density for convergence of the results.

For the tests using varied fitting sets, the results are summarised in \cref{tab:chi2full}, where the $\chi^2$ values calculated for the HLM and LCM obtained with \basta are compared to the corresponding $\chi^2$ value for \bm{12} and \bm{13}. The determined \tcore from \basta is reported alongside the $\chi^2$ values, for comparison with the \tcore obtained in \cref{sec:results}.

For fitting of non-asteroseismic parameters the determined values listed in \cref{subsec:observations} are used. The results in \cref{sec:results} were obtained by fitting the set $(\feh,\teff,\rho)$, while the forward modelling approach of \buld did minimisation in regards to the set $\left(\feh, \teff, \lphot, \log g, \rho\right)$. We explore the implications of using partly and fully the same set as \buld to determine their influence on the inferred \tcore, using the standard \basta fitting set $(\feh,\teff)$ as a baseline. Additionally, it is tested how adding a fit of the distance through the parallax and apparent magnitudes affects the result. As \bm{12} and \bm{13} does not include synthetic magnitudes, the distance have not been fitted to these models and their $\chi^2$ values are thus not reported for this case.

The full set of possible ratio fitting choices as described in \cref{subsec:chi2r} has been repeated for the standard $\left(\feh, \teff\right)$ fitting set, cases 1-16 in \cref{tab:chi2full}. For the remaining fitting sets, the recommended ratio fitting case as used in \cref{sec:results} (case 16 in \cref{tab:chi2r}) and the ratio fitting case with $r_{012}$ but opposite choices of the remaining ratio fitting variations (case 9 in \cref{tab:chi2r}) are used throughout cases 17-30, 33 and 34 in \cref{tab:chi2full}. Case 31 fits the $r_{02}$ sequence as was done in \buld along with their chosen classical parameters, while case 32 does the same, but with our recommendation of removing the flagged modes and including the covariance. 

Overall, the LCM has the lowest $\chi^2$ across all cases. This is primarily caused by the contribution from \teff being higher for \bm{12} and \bm{13}. Comparing the $\chi^2$ of these two models, only case 31 leads to \bm{13} having a lower $\chi^2$ than \bm{12}. One would expect this to also be the case for fitting case 5 and 7, as the ratios for these cases were determined to favour \bm{13} in \cref{subsec:chi2r}. However, due to the normalisation of $\chi_r^2$, the total $\chi^2$ is dominated by the contribution from \teff, which deviates more for \bm{13}. The case of not normalising the ratios is discussed at the end of this section.

As can be seen across all fitting cases in \cref{tab:chi2full}, the inferred median \tcore is stable at $\sim \SI{0.6}{\giga\year}$ for fits using the $r_{01}$ ratios. The $r_{02}$ ratios alone shows to have a poor inference power on this parameter, as expected since it has been shown to carry little information on the location of the convective zones \citep{Roxburgh2009}. The upper bound (84th quantile of the posterior) vary more across fitting sets, but generally places an upper limit at $\SI{\sim3.5}{\giga\year}$. As was mentioned in \cref{subsec:best_fit_models} this large span in inferred values imply that the ratios most likely do not provide a direct constraint on \tcore. They however provide the best constraint on internal structure, and is thus the best available parameter to attempt to constrain \tcore.

Case 31 shows a preference of \bm{13} over \bm{12} when fitting as was done in \buld, recovering their solution. Case 32 shows that by removing the flagged modes and taking covariances into account, the opposite preference of models are obtained. The determined 84th percentile on \tcore for both cases are twice as large as the cases fitting $r_{012}$, showing that this choice of fitting parameters provide a poor constraint on \tcore.

To verify the results using different photometric observations than those used in \buld, the standard fitting case was repeated including the fitting of distance/parallax in cases 33 and 34 \citep[method described in section 4.2.2 of][]{BASTA}. It uses the Tycho-2 magnitudes, as mentioned in \cref{subsec:observations}, as the Gaia-magnitudes were found not to converge at a solution within the grid. The results from these cases are consistent with the other cases, and thus the determined \tcore seems independent on the choice of photometric observables.

It is worth noting that the HLM and LCM models from the \basta fits often have $\tcore = 2-\SI{5}{\giga\year}$. As described in \cref{sec:results,fig:profile_differences}, these are found to be older models where the signature of the convective core has almost disappeared from the profiles of the internal structure. While not immediately apparent in the determined ages (e.g.\@ as given in \cref{fig:corner}), this could lead to a systematic increase in the inferred values. It should however still be noted that the models are obtained with parameters and physics outside our current understanding of appropriate values. The ratios can reliably determine that \kepl does not have a convective core at its current age. However, models with induced early convective cores lead to drive the fit to older models, where the influence of the convective core has disappeared from the model, and due to the ratios providing the best constraint on the age of the star, this is not counteracted by other observables. The age of the model is however also dependent on the input physics and parameters, and this effect can therefore not be studied in detail here. 

It is important to note that this trend is not observed when comparing \bm{12} and \bm{13}. Here, \bm{13} with a large \tcore has a lower inferred age than \bm{12}, while the difference in age is also lower than what is observed for the models in this work. Thus, this effect might be mostly related to the way overshooting is prescribed in the models presented in this work. However, to definitively determine this, it would require a larger set of models using the same framework as \buld to compare to, and is thus not explored further in this work.

When comparing cases 17 through 24 to case 9 and 16, it becomes apparent that while $\rho$ and $\log g$ increases the $\chi^2$ of \bm{12} and \bm{13} almost equally, \lphot increases the $\chi^2$ of \bm{12} significantly more than for \bm{13}. This difference is larger than the discrepancy between the $\chi^2$ of the models in case 31 where \bm{13} is preferred. The conclusion to favour \bm{13} over \bm{12} is thus in terms of classical parameters primarily driven by the fitted \lphot value. Meanwhile, this does not increase the determined value of \tcore in the fits using \basta, showing no correlation between this parameter and the convective core lifetime of the models in this work. 

Finally, we also repeated the analysis outlined in this subsection with each frequency ratio weighted as an individual measurement, as mentioned at the start of the section (i.e. setting $N_i$ to 1 in \cref{eq:chi2} when computing $\chi^2_r$). Across all cases, it leads to a large reduction in the number of models with high likelihoods, which causes the posteriors (\cref{fig:corner}) to become more jagged and multi-modal. However, the convective core size evolution of the highest likelihood tracks presented in \cref{fig:core_best} and the overall determined parameters does not change significantly. Specifically, \tcore changes to have a median value around \SI{1}{\giga\year}, with an upper bound around \SI{\sim4.5}{\giga\year} (compared with the \SI{\sim0.6}{\giga\year} and \SI{\sim3.5}{\giga\year} from fitting with $N_i$ equal to the number of ratios), with the increase heavily driven by two tracks in the posterior with a large likelihood, that has high \tcore and ages. Specifically, these ages are \SI{\sim13}{\giga\year}, above the 84th percentile of the posterior, and close to the age of the Universe. The inferred larger upper bound using this unweighted scheme is therefore correlated with an increased inferred age.

In terms of the comparison between \bm{12} and \bm{13} when the frequency ratios are weighted as individual measurements, it shows a lower $\chi^2$ for \bm{13} than \bm{12} in cases 5, 7 and 31. This change for case 5 and 7 is due to the deviation of \teff for \bm{13} not dominating the $\chi^2$, as is the case when the ratio contribution is normalised. The conclusion of case 31 remains unchanged, as the $\chi^2$ contribution of \lphot already compensated for the \teff  contribution.

From the discussion above, and the fits presented in \cref{tab:chi2full}, we can conclude that \tcore can be constrained to be lower than \SI{4}{\giga\year} as this is consistent with all fits apart from the $r_{02}$ ratio fits. As mentioned, this is obtained using non-standard stellar models, where the expected value would be constrained to be within \SI{1}{\giga\year}.

\section{Conclusions}
\label{sec:conclusion}
In this work, we have studied the survival of the initial convective core of \kepl. In the work of \cite{Buldgen2019} it was determined to have an unusually long lifetime of the convective core $(\SI{\sim7.85}{\giga\year})$ of the old $(\SI{\sim11\pm1}{\giga\year})$ sub-solar mass $(\SI{\sim0.748\pm0.044}{\solarmass})$ star, i.e.\@ an order of magnitude longer than what our current understanding of stellar evolution suggests. We produced a grid of stellar models with a high sampling of input parameters and artificially induced long convective-core lifetimes to determine the properties of \kepl by fitting the observations to the grid using the \basta code. The determined fundamental properties were consistent with previous studies, however we found a convective-core lifetime of $0.597^{+2.151}_{-0.244}\,{\rm Gyr}$ by fitting the effective temperature, \teff, metallicity, \feh, mean density, $\rho$, and the $r_{012}$ frequency separation ratio sequence to the grid. 

While this lifetime is reasonably large, inspecting the highest-likelihood models in the grid reveals this to possibly not be due to a signature in the observed frequency ratios. Rather, the models with a high likelihood have no convective core signature left in the interior profile, where the models with a longer convective-core lifetime have evolved for longer before reaching this state. This indicates that the inclusion of an artificially induced longer convective-core lifetime introduces an age bias, systematically increasing the determined age of the star, when using the framework for stellar models applied in this study.

Additionally, we varied the sets of observations used when fitting to further study what can lead to the inference of long convective-core lifetimes. First, the choices related to how the mode frequency ratios are fitted to the best-fit models from \cite{Buldgen2019} were varied. The largest differences were seen when switching between including or excluding two potentially unreliable frequency ratios flagged by \cite{Davies2016} (failing their quality check) from the fitted ratios. This is also in agreement with the possibility for high-frequency ratios being shifted to larger amplitudes due to surface magnetic activity, making them less reliable \citep{Thomas2021}. Large differences were also seen when including or excluding the covariances between the ratios in the computation of the error function. As the frequency ratios per construction repeatedly use the same frequency modes in their determination, they are inherently correlated and these correlations are important to consider. The conclusion to prefer a model with a long convective-core lifetime was only recovered when the two prior-dominated frequencies were included, when the covariance of the mode frequency ratios was not considered, or when only fitting the $r_{02}$ frequency ratios. 

Lastly, we repeated the fit for different sets of classical parameters to determine what observations drive the inference of a long convective-core lifetime and to verify the stability of the determined convective-core lifetime across the observed parameters. Only fits using the frequency ratio fit choices as mentioned above and including the luminosity derived by \cite{Buldgen2019} as a fitting parameter recovered the conclusion of preferring a long lifetime convective core. From the Bayesian marginalised posteriors determined from each fit, the median and 84th percentile were extracted with the median reliably around \SI{0.6}{\giga\year}, while the 84th percentile was reliably within \SI{\sim3.5}{\giga\year}. This large upper bound on convective core lifetime does however suggest that frequency ratios presumably only provide a weak constraint on this parameter.

This study exemplifies the difficulties of accurately inferring stellar interiors, both in the present and how it evolves. While the approach of varying the geometric cut-off \gc in the overshoot description allowed for varied sizes and evolution of convective cores, there are no physical argument for changing its value from the solar-calibrated one. The change of inferred core evolution with inclusion or exclusion of the two frequency modes highlight the difficulties in determination of mode frequencies. While they can be determined to a high precision, the heavily prior-reliant methods for their determination can not always guarantee their accuracy. Also, if one does not consider the covariance of the mode frequency ratios it can easily lead to wrong conclusions as the differences in ratios between different models can be quite minute. Lastly, the appearance of an early convective core might not be well constrained by the frequency ratios and it in some cases simply leads to determination of older models where the influence of the early convective core had almost disappeared.

This is therefore not conclusive proof that \kepl did not have a large convective core during its early evolution, but the most likely scenario is that it did not. Additionally, the non-zero determination could simply be the result of the convective-core lifetime producing older models as they take longer to reach an interior structure without signatures of a convective core. However, we can conclude that the model from \cite{Buldgen2019} without a long convective-core lifetime is preferred over their originally determined best fit model with a long convective-core lifetime within our choices of frequency ratios fitting. Our analysis reliably determines a short convective-core lifetime with determined stellar properties consistent with previous studies, and with standard stellar evolution theory.

\section*{Acknowledgements}
The authors thank Dr.~Gaël~Buldgen for the cooperation during this work, for providing the data needed for the comparisons, and for the very helpful and detailed comments as referee of this work. The authors also thank Prof.~Dr.~Jørgen~Christensen-Dalsgaard and Prof.~Dr.~Achim~Weiss for contributing their expertise on stellar evolution models for this work. 

MLW acknowledges support from the Carlsberg Foundation (grant agreement CF19-0649). Funding for the Stellar Astrophysics Centre is provided by The Danish National Research Foundation (Grant agreement No.~DNRF106). AS acknowledges support from the European Research Council Consolidator Grant funding scheme (project ASTEROCHRONOMETRY, G.A. n. 772293, \url{http://www.asterochronometry.eu}). The numerical  results presented in this work were obtained at the Centre for Scientific Computing, Aarhus \url{http://phys.au.dk/forskning/cscaa/}. This work has made use of data from the European Space Agency (ESA) mission {\it Gaia} (\url{https://www.cosmos.esa.int/gaia}), processed by the {\it Gaia} Data Processing and Analysis Consortium (DPAC, \url{https://www.cosmos.esa.int/web/gaia/dpac/consortium}). Funding for the DPAC has been provided by national institutions, in particular the institutions participating in the {\it Gaia} Multilateral Agreement.

\section*{Data Availability}
The data and code for the data presented in this paper is available at \url{https://github.com/MLWinther/Kepler444}.



\bibliographystyle{mnras}
\bibliography{bibliography} 




\appendix
\section{Equilibrium of $^3$He}
\label{app:he3}

The equilibrium mass fraction of \ce{^3He}, $\left(X_{\ce{^3He}}\right)_{\rm e}$, present in the core can be derived following \cite{Iliadis2015}. It is based on the reaction rates in the core and the presence of other materials, where the $S$-factors used for the reaction rates are those of \cite{Sfactor}. This assumes an isolated system with no material entering from exterior sources. As the consumption rate of \ce{^3He} is larger than its creation rate in the PP-chains dominating the reactions in the core, there should be a low amount of \ce{^3He}, only present in the delay between creation and consumption. The theoretical equilibrium amount can then be computed for the core evolution of a stellar model and compared to the actual amount present, to determine if any has been fed to the core through mixing with the surrounding material. 

\begin{figure}
    \centering
    \includegraphics[width=\linewidth]{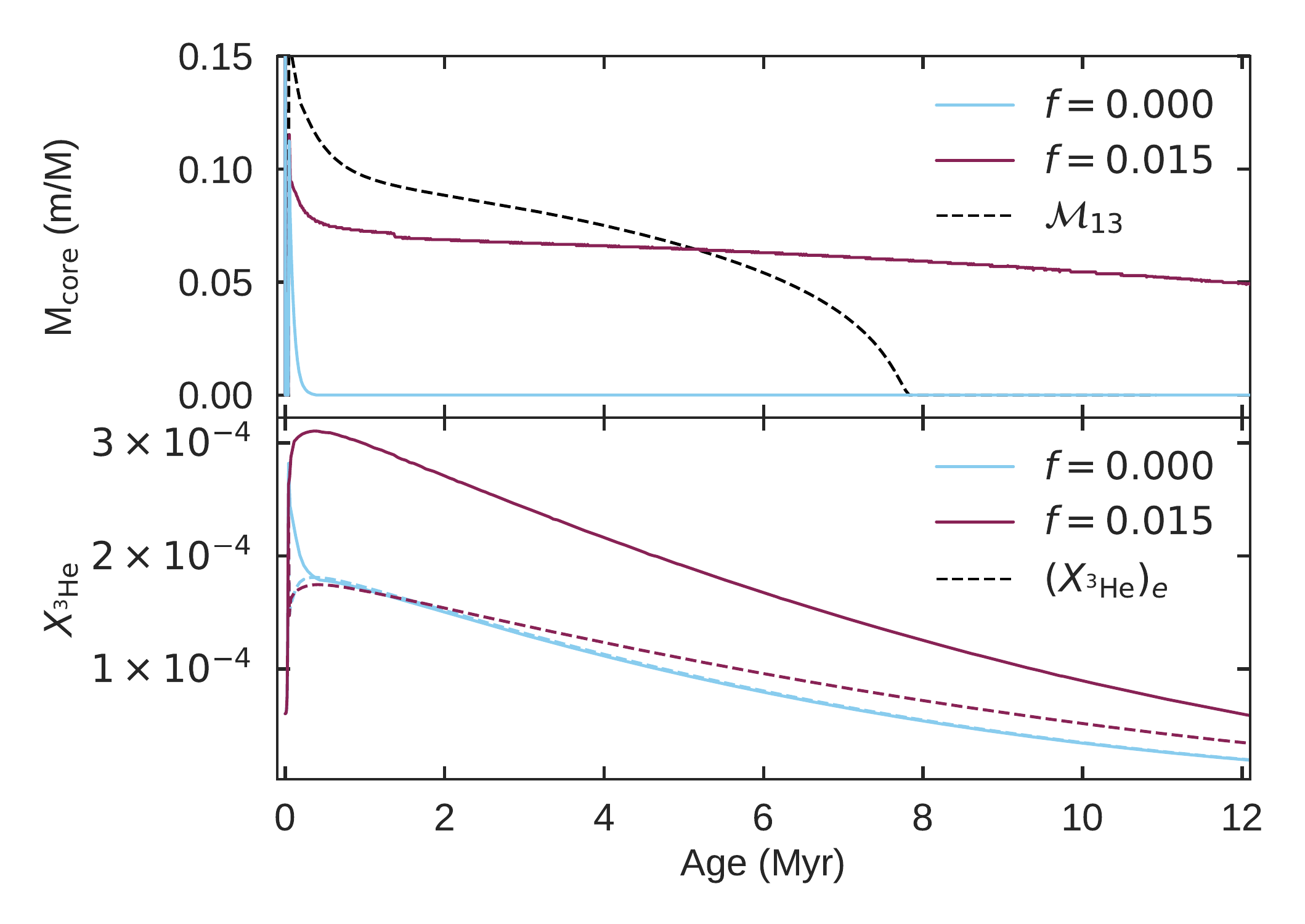}
    \caption{The evolution of the convective core size (\textit{left}) and \ce{^3He} abundance (\textit{right}) as a function of age, of two \textsc{Garstec} models of similar input to that of \bm{13} apart from overshooting. The model with a large overshooting efficiency clearly shows an increased abundance of \ce{^3He} compared to the model without overshooting, which is also a large increase compared to the equilibrium amount.}
    \label{fig:he3_equilibrium}
\end{figure}

An example of this can be seen in \cref{fig:he3_equilibrium}, where the convective core mass fraction and central \ce{^3He} mass fraction evolution is shown for two \textsc{Garstec} models with the same input as \bm{13}, for two different values of overshooting efficiency. For $f_{\rm ove}=0$, and thereby no overshoot, the abundance of \ce{^3He} follows the computed equilibrium, while for $f_{\rm ove}=0.015$, it is clearly larger ($\approx 1.8$ times as much, slowly reducing) whereby it shows that overshooting indeed injects additional \ce{^3He} into the core. The abundance evolution is almost identical to those presented in \cref{fig:buld_profile} before an age of \SI{\sim 7.85}{\giga\year}, whereby one can conclude that \bm{12} follows the equilibrium amount of \ce{^3He}, while \bm{13} only does so towards the end of its evolution, after the convective core disappears. 

\section{Échelle of best-fit models}
\label{app:echelle}
\begin{figure}
    \centering
    \includegraphics[width=\linewidth]{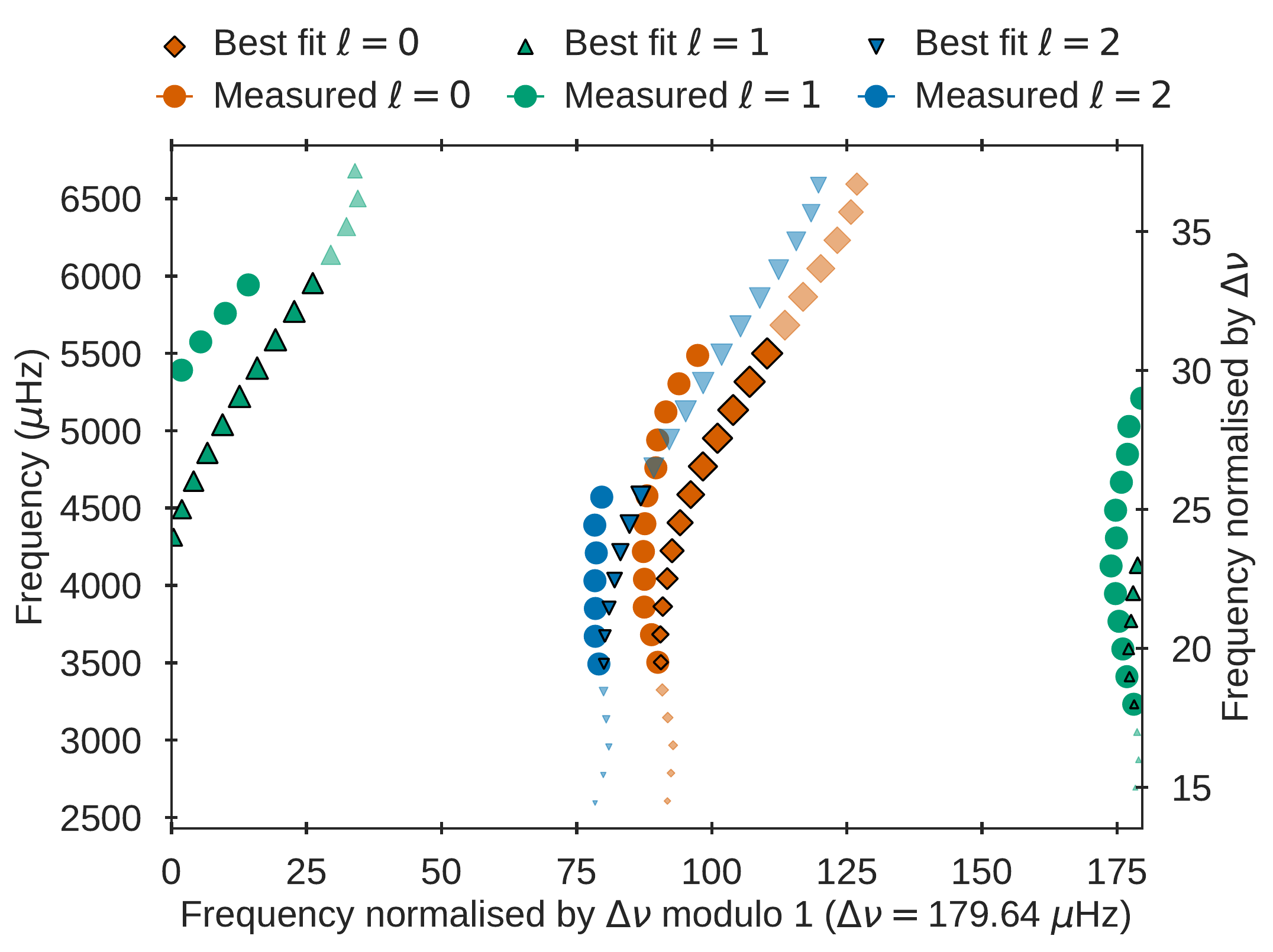}
    \caption{Échelle diagram of the observed frequencies and the uncorrected model frequencies of the HLM. The two prior dominated frequencies $\nu\left(\ell=2,n=24,25\right)$ have been omitted.}
    \label{fig:ech_uncorrect}
\end{figure}

\begin{figure}
    \centering
    \includegraphics[width=\linewidth]{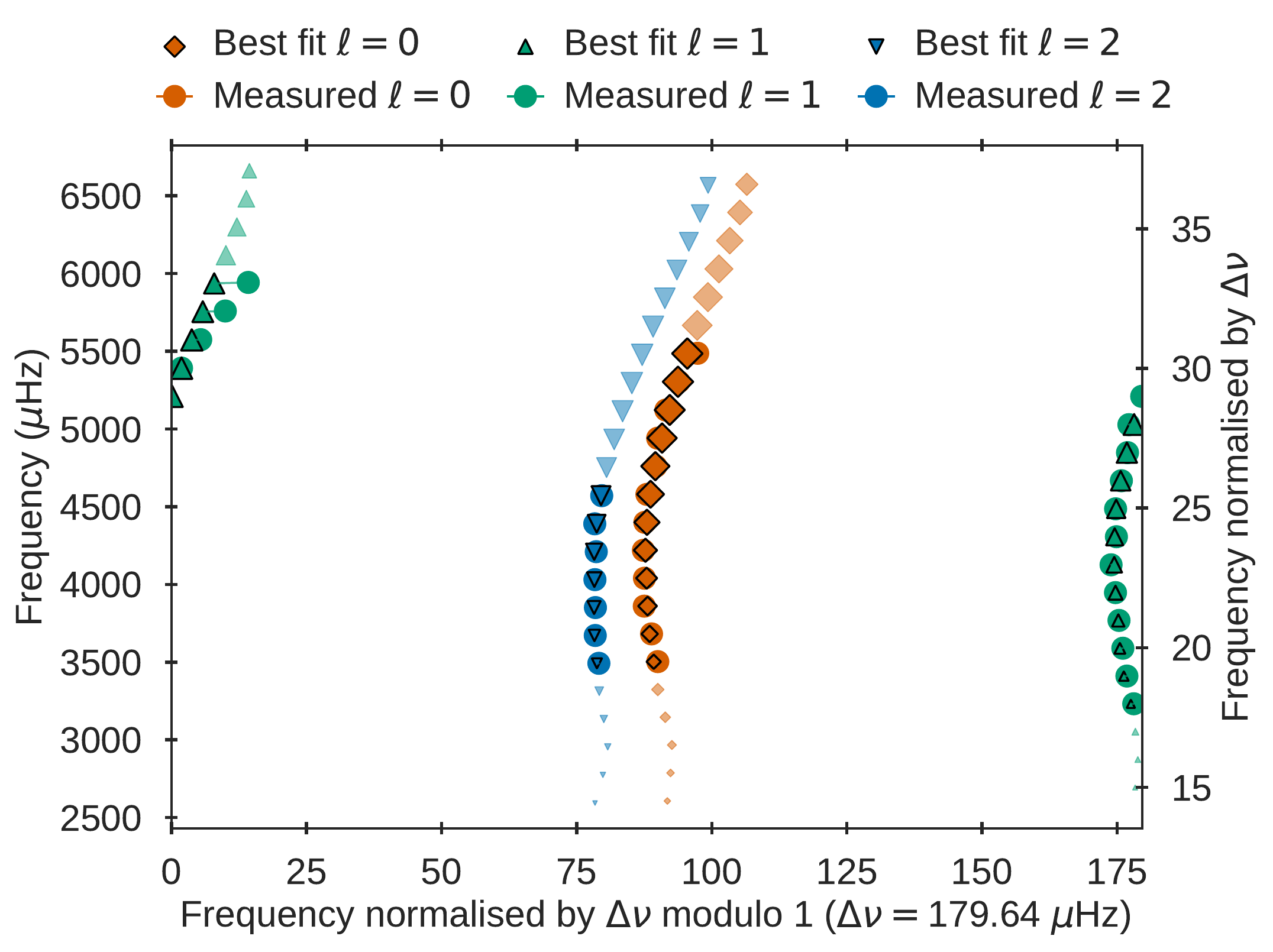}
    \caption{Échelle diagram of the observed frequencies and the model frequencies of the HLM corrected using the combined correction of \protect\cite{BG14}. The two prior dominated frequencies $\nu\left(\ell=2,n=24,25\right)$ have been omitted.}
    \label{fig:ech_correct}
\end{figure}

While the frequency ratios of the HLM are shown in \cref{fig:ratios}, it can also be informative to look at the échelle diagrams. The uncorrected frequencies are shown in \cref{fig:ech_uncorrect}, while a version with model frequencies corrected for the surface effect is shown in \cref{fig:ech_correct}. Here the two term correction from \cite{BG14} were used with the fitted coefficients $a_{-1} = \num{7.308e-10}$ and $a_{3} = \num{-7.331e-09}$.

The large offset between modelled and observed frequencies in the uncorrected échelle diagram \cref{fig:ech_uncorrect} indicates that the HLM is slightly too young.

\section{Grid resolution}
\label{app:grid}
\begin{figure*}
    \centering
    \includegraphics[width=\linewidth]{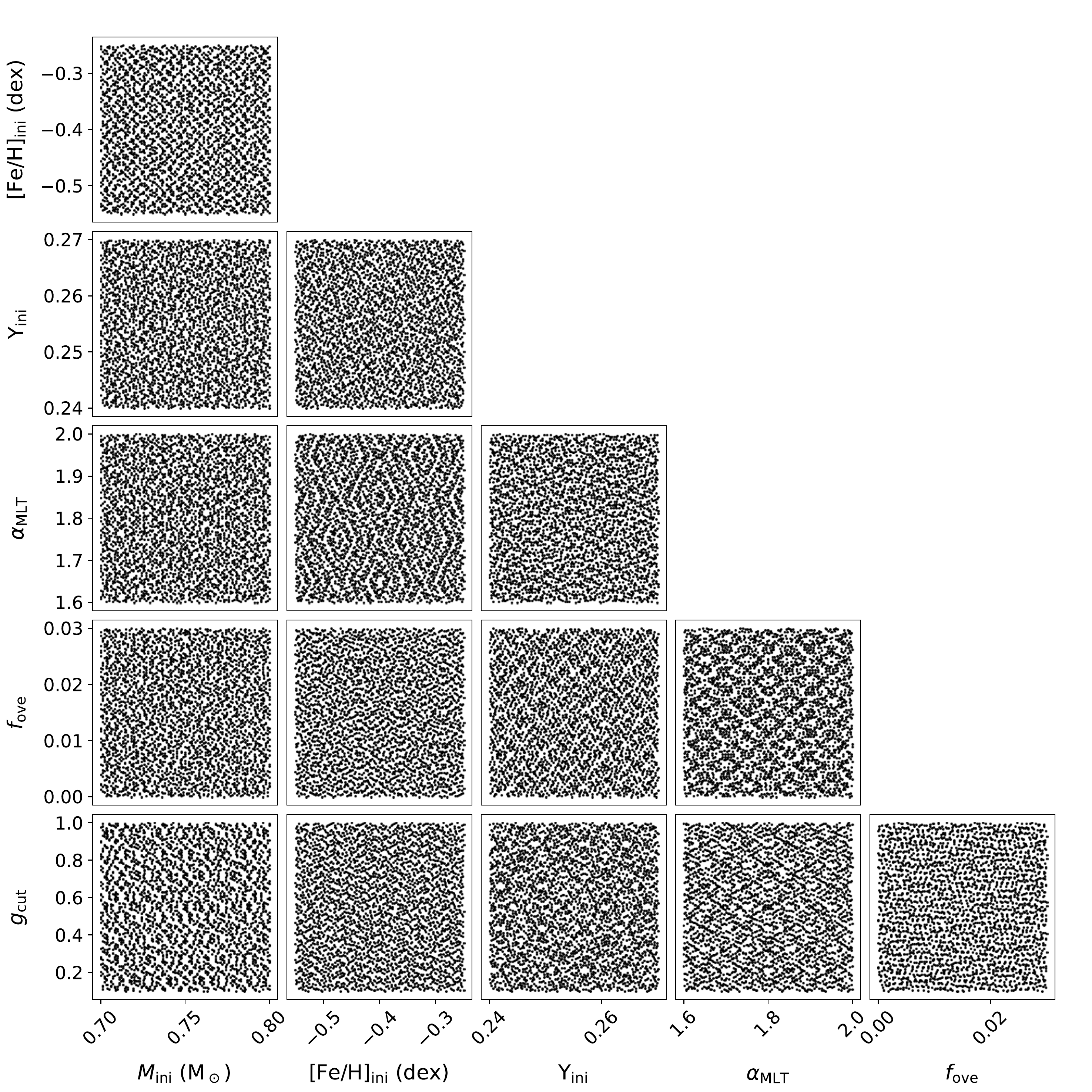}
    \caption{Distribution of initial parameters of tracks in the grid. $\left[ \alpha/\mathrm{Fe}\right]$ is not shown, as it can only be one of two discrete values.}
    \label{fig:grid_coverage}
\end{figure*}

The parameters for each track in the grid described in \cref{subsec:kepl_grid} are drawn from continuous distributions, using the (quasi-)random Sobol-sequence \citep{Sobol1,Sobol2,Sobol3,Sobol4,Sobol5,Sobol6}, as described in \cref{sec:modelling}. This ensures an uniform sampling of the parameter space, and avoids clumpy over-densities in the parameter space that is usually encountered for Cartesian or pseudo-random methods \citep{BASTA}. Only two parameters are discretised, initial mass as the starting models have only been prepared in steps of \SI{0.001}{\solarmass}, and alpha-element enhancement as opacity tables have only been computed in steps of \SI{0.1}{\dex}.

Due to the grid being constructed this way, it does not have a regular measure of resolution in each parameter, as is e.g. the case for Cartesian-sampled grids with a fixed sampling in each parameter. To represent the resolution of the grid, the initial parameters of the track that constitute the grid are instead shown in \cref{fig:grid_coverage} for each combination of parameters. This shows a high sampling of each combination of parameters, while some slight repetition of patterns within these also shows the sampling to not be completely random.



\bsp	
\label{lastpage}
\end{document}